\documentclass{siamltex}
\input epsf

\def\Afrak{{\bf A}}
\def\Bfrak{{\bf B}}

\def\Hess{\mathop{\rm Hess}\nolimits}
\def\sym{\mathop{\rm sym}\nolimits}
\def\skew{\mathop{\rm skew}\nolimits}
\def\matsep{,\mskip 6mu plus 2mu\relax}
\def\half{{\textstyle{1\over2}}}
\def\by{\hbox{-by-}}
\newcommand{\PS}{{\bf P}}
\newcommand{\R}{{\bf R}}

\newcommand{\ha}{\frac{1}{2}}
\newcommand{\tr}{\mathop{\rm tr}\nolimits}
\newcommand{\T}{^T\!}
\newcommand{\Yd}{{\dot Y}}
\newcommand{\Ydd}{{\ddot Y}}
\newcommand{\Yperp}{{Y_{\!\perp}}}
\newcommand{\Qd}{{\dot Q}}

\def\squarebox#1{\hbox to #1{\hfill\vbox to #1{\vfill}}}

\def\cl{{\bf L}}

\def\pldx{{{{\bf L}}_x}}
\def\pldxx{{{{\bf L}}_{xx}}}
\def\pldxxk{{{{\bf L}}^k_{xx}}}

\newdimen\sqbox \sqbox=48pt
\newdimen\rcbox \rcbox=12pt
\newlength{\boxwidth}
\newcommand{\boxer}[1]{\begin{center} \fbox{\fbox{
     \begin{minipage}{\boxwidth} #1 \end{minipage}}}
\end{center}}

\def\boxit#1{\vbox{\hrule\hbox{\vrule\kern0pt
      \vbox{\kern0pt\hbox{#1}\kern0pt}\kern0pt\vrule}\hrule}}
\def\boxitop#1{\vtop{\hrule\hbox{\vrule\kern0pt
     \vbox{\kern0pt\hbox{#1}\kern0pt}\kern0pt\vrule}\hrule}}
\def\boxedQ{\vcenter{\boxit{\rule[-.4\sqbox]{0pt}{\sqbox}%
  \hbox to\sqbox{\hss$Q$\hss}}}}
\def\boxedY{\vcenter{\boxit{\rule[-.4\sqbox]{0pt}{\sqbox}%
  \hbox to\rcbox{\hss$Y$\hss}}}}

\newcommand{\masbox}[1]{\begin{center} \vspace{0.1in}
  {\framebox[\hsize][c]{
     \begin{minipage}{\boxwidth} \vspace{0.1in} #1 \vspace{0.1in}
   \end{minipage}}} \vspace{0.1in} \end{center}}

\newcommand{\stepitem}{\par\vskip6pt\noindent\hskip2em\hangindent2em\hangafter1
  \llap{$\bullet$\enspace}\ignorespaces}
\newcommand{\foritem}{\par\vskip3pt\noindent\hskip4em\hangindent4em\hangafter1
  \llap{$\circ$\enspace}\ignorespaces}

\makeatletter
\def\rmatrix#1{\left(\null\,\vcenter{\scriptsize\normalbaselines\m@th
    \ialign{\hfil$\checkminus ##$&&\,\,\hfil$\checkminus ##$\crcr
      \mathstrut\crcr\noalign{\kern-\baselineskip}
      #1\crcr\mathstrut\crcr\noalign{\kern-\baselineskip}}}\,\right)}
\def\checkminus{\futurelet\next\ch@ckminus}
\def\ch@ckminus{\ifx\next-\def\temp{}\else\def\temp{\hphantom-}\fi\temp}
\def\eqalign#1{\null\,\vcenter{\openup\jot\m@th
  \ialign{\strut\hfil$\displaystyle{##}$&$\displaystyle{{}##}$\hfil
      \crcr#1\crcr}}\,}

{\catcode`\@=11\obeyspaces\obeylines
\gdef\beginmatlab{\begingroup\scriptsize\baselineskip=10pt\tt\frenchspacing %
  \spaceskip=\z@ \xspaceskip=\z@ %
  \parskip=\z@ \parindent=2em
  \chardef\other=12  %
  \catcode`\"=\z@    %
  \catcode`\&=\other %
  \catcode`\^=\other %
  \catcode`\_=\other %
  \catcode`\~=\other %
  \catcode`\%=\other %
  \catcode`\\=\other %
  \obeyspaces\gdef {\ifvmode\leavevmode\fi\nobreak\space}%
  \obeylines\gdef^^M{\ifvmode\penalty-100\vskip6pt\fi\par\penalty9999}}}

\makeatother

\title{The Geometry of Algorithms\\with Orthogonality
  Constraints}

\author{
Alan Edelman\thanks{
Department of Mathematics Room 2-380,
Massachusetts Institute of Technology,
Cambridge, MA 02139,
edelman@math.mit.edu,
http://www-math.mit.edu/\char`\~edelman.
Supported by a fellowship from the Alfred P. Sloan Foundation
and NSF grants 9501278-DMS and 9404326-CCR.
}\and
T.A. Arias\thanks{
Department of Physics,
Massachusetts Institute of Technology,
Cambridge, MA 02139,
\hbox{muchomas@mit.edu}.
Supported by an NSF/MRSEC Seed Project grant from the MIT Center for
Material Science and Engineering.
}
\and
Steven T. Smith\thanks{
MIT Lincoln Laboratory,
244 Wood Street, 
Lexington, MA 02173,
{stsmith@ll.mit.edu}.
Sponsored by DARPA under Air~Force contract
F19628-95-C-0002. Opinions, interpretations, conclusions, and
recommendations are those of the author and are not necessarily
endorsed by the United States Air~Force.
}}
\begin{document}

\maketitle

\newif\ifDRAFT \DRAFTfalse
\ifDRAFT
  \nointerlineskip
  \vbox to0pt{\kern-1.25in\leftline{\footnotesize {\sc SIAM J.\ Matrix Anal.\
    Appl.}, to appear.\hfil}\vss}
  \nointerlineskip
\fi

\begin{center}
\underline{In press, {\em SIAM J.\ Matrix Anal.\ Appl.}}
\end{center}
\vspace{0.2in}

\begin{abstract}
In this paper we develop new Newton and conjugate gradient algorithms
on the Grassmann and Stiefel manifolds.  These manifolds represent the
constraints that arise in such areas as the symmetric eigenvalue
problem, nonlinear eigenvalue problems, electronic structures
computations, and signal processing.  In addition to the new
algorithms, we show how the geometrical framework gives penetrating
new insights allowing us to create, understand, and compare
algorithms.  The theory proposed here provides a taxonomy for
numerical linear algebra algorithms that provide a top level
mathematical view of previously unrelated algorithms.  It is our hope
that developers of new algorithms and perturbation theories will
benefit from the theory, methods, and examples in this paper.
\end{abstract}

\begin{keywords} 
  conjugate gradient, Newton's method, orthogonality constraints,
  Grassmann manifold, Stiefel manifold, eigenvalues and
  eigenvectors, invariant subspace, Rayleigh quotient iteration,
  eigenvalue optimization, sequential quadratic programming,
  reduced gradient method, electronic structures computation,
  subspace tracking
\end{keywords}

\begin{AMS}
49M07, 49M15, 53B20,
65F15, 15A18, 51F20, 81V55
\end{AMS}

\pagestyle{myheadings}
\thispagestyle{plain}

\markboth{Edelman, Arias, and Smith}{Orthogonality Constraints}

\section{Introduction}

Problems on the Stiefel and Grassmann manifolds arise with
sufficient frequency that a unifying investigation of algorithms
designed to solve these problems is warranted.  Understanding
these manifolds, which represent orthogonality constraints (as in
the symmetric eigenvalue problem), yields penetrating insight
into many numerical algorithms and unifies seemingly unrelated
ideas from different areas.

The optimization community has long recognized that linear and
quadratic constraints have special structure that can be
exploited.  The Stiefel and Grassmann manifolds also represent
special constraints.  The main contribution of this paper is a
framework for algorithms involving these constraints, which draws
upon ideas from numerical linear algebra, optimization,
differential geometry, and has been inspired by certain problems
posed in engineering, physics, and chemistry.  Though we do
review the necessary background for our intended audience, this
is not a survey paper. This paper uses mathematics as a tool so
that we can understand the deeper geometrical structure
underlying algorithms.

In our first concrete problem we minimize a function $F(Y)$,
where $Y$ is constrained to the set of $n\by p$ matrices such
that $Y^TY=I$ (we call such matrices orthonormal), and we make
the further homogeneity assumption that $F(Y)=F(YQ)$, where $Q$
is any $p\by p$ orthogonal matrix.  In other words, the objective
function depends only on the subspace spanned by the columns of
$Y$; it is invariant to any choice of basis.  The set of
$p$-dimensional subspaces in~$\R^n$ is called the Grassmann
manifold.  (Grassmann originally developed the idea in 1848, but
his writing style was considered so obscure \cite{bios} that it
was appreciated only many years later.  One can find something of
the original definition in his later work
\cite[Chap.~3, Sec.~1,Article~65]{grassmann}.)  To the best of our
knowledge, the geometry of the Grassmann manifold has never been
explored in the context of optimization algorithms, invariant
subspace computations, physics computations, or subspace
tracking.  Useful ideas from these areas, however, may be put
into the geometrical framework developed in this paper.

In our second problem we minimize $F(Y)$ without the homogeneity
condition $F(Y)=F(YQ)$ mentioned above, i.e., the optimization
problem is defined on the set of $n\by p$ orthonormal matrices.
This constraint surface is known as the Stiefel manifold, which
is named for Eduard Stiefel, who considered its topology in the
1930s \cite{stiefel35}.  This is the same Stiefel who in
collaboration with Magnus Hestenes in 1952 originated the
conjugate gradient algorithm \cite{hestenes52}.  Both Stiefel's
manifold and his conjugate gradient algorithm play an important
role in this paper.  The geometry of the Stiefel manifold in the
context of optimization problems and subspace tracking was
explored by Smith \cite{smith93a}. In this paper we use numerical
linear algebra techniques to simplify the ideas and algorithms
presented there so that the differential geometric ideas seem
natural and illuminating to the numerical linear algebra and
optimization communities.

The first author's original motivation for studying this problem
came from a response to a linear algebra survey \cite{edelman93},
which claimed to be using conjugate gradient to solve large dense
eigenvalue problems.  The second and third authors were motivated
by two distinct engineering and physics applications.  The
salient question became: ``What does it mean to use conjugate
gradient to solve eigenvalue problems?''  Is this the Lanczos
method?  As we shall describe, there are dozens of proposed
variations on the conjugate gradient and Newton methods for
eigenvalue and related problems, none of which are Lanczos.
These algorithms are not all obviously related.  The connections
among these algorithms have apparently not been appreciated in
the literature while in some cases numerical experiments have
been the only basis for comparison when no theoretical
understanding was available.  The existence of so many variations
in so many applications compelled us to ask for the big picture:
What is the mathematics that unifies all of these apparently
distinct algorithms.  This paper contains our proposed
unification.

We summarize by itemizing what is new in this paper.
\begin{remunerate} 

\item Algorithms for Newton and conjugate gradient methods on the
      Grassmann and Stiefel manifolds that naturally use the geometry
      of these manifolds.  In the special cases that we are aware of,
      our general algorithms are competitive up to small constant
      factors with the best known special algorithms.  Conjugate
      gradient and Newton on the Grassmann manifold have never been
      studied before explicitly.  Stiefel algorithms have been studied
      before \cite{smith93a}, but the ideas here represent
      considerable simplifications.

\item A geometrical framework with the right mix of abstraction and
      concreteness to serve as a foundation for any numerical
      computation or algorithmic formulation involving orthogonality
      constraints, including the symmetric eigenvalue problem.  We
      believe that this is a useful framework because it connects
      apparently unrelated ideas; it is simple and mathematically
      natural. The framework provides new insights into existing
      algorithms in numerical linear algebra, optimization, signal
      processing, and electronic structures computations, and it
      suggests new algorithms.  For example, we connect the ideas of
      geodesics and the cubic convergence of the Rayleigh quotient
      iteration, the CS decomposition, and sequential quadratic
      programming.  We also interpret the ill-conditioning of
      eigenvectors of a symmetric matrix with multiple eigenvalues as
      the singularity of Stiefel and Grassmann coordinates.

\item Though geometrical descriptions of the Grassmann and Stiefel
      manifolds are available in many references, ours is the first to
      use methods from numerical linear algebra emphasizing
      computational efficiency of algorithms rather than abstract
      general settings.

\end{remunerate} 

The remainder of this paper is organized into three sections.
The geometrical ideas are developed in
\S\ref{sec:diffgeom}.  This \S\ref{sec:unified}
provides a self-contained introduction to geometry, which may not
be familiar to some readers, while deriving the new geometrical
formulas necessary for the algorithms of
\S\ref{sec:geomalg} and the insights of
\S\ref{sec:geomalg} provides descriptions of new algorithms
for optimization on the Grassmann and Stiefel manifolds.
Concrete examples of the new insights gained from this point of
view are presented in \S\ref{sec:unified}.  Because we wish
to discuss related literature in the context developed in
\S{}\ref{sec:diffgeom} and \S\ref{sec:geomalg}, we defer
discussion of the literature to \S\ref{sec:unified} where
specific applications of our theory are organized.

\section{Differential Geometric Foundation for Numerical Linear Algebra}
\label{sec:diffgeom}

A geometrical treatment of the Stiefel and Grassmann manifolds
appropriate for numerical linear algebra cannot be found in
standard differential geometry references.  For example, what is
typically required for practical conjugate gradient computations
involving $n\by p$ orthonormal matrices are algorithms with
complexity of order~$np^2$.  In this section we derive new
formulas that may be used in algorithms of this complexity in
terms of standard operations from numerical linear algebra.
These formulas will be used in the algorithms presented in the
following section.  Because we focus on computations, our
approach differs from the more general (and powerful)
coordinate-free methods used by modern geometers
\cite{chavel93,helgason78,kobayashi,oneill83,spivak79,warner83}.
Boothby \cite{boothby86} provides an undergraduate level
introduction to the coordinate-free approach.

For readers with a background in differential geometry, we wish
to point out how we use extrinsic coordinates in a somewhat
unusual way.  Typically, one uses a parameterization of the
manifold (e.g., $x=\cos u\sin v$, $y=\sin u\sin v $, $z=\cos v$
for the sphere) to derive metric coefficients and Christoffel
symbols in terms of the parameters ($u$ and $v$).  Instead, we
only use extrinsic coordinates subject to constraints (e.g.,
$(x,y,z)$ such that $x^2+y^2+z^2=1$).  This represents points
with more parameters than are intrinsically necessary, but we
have found that the simplest (hence computationally most useful)
formulas for the metric and Christoffel symbol are obtained in
this manner.  The choice of coordinates does not matter
abstractly, but on a computer the correct choice is essential.

We now outline this section.  After defining the manifolds of
interest to us in \S\ref{sec:def}, we take a close look at
the Stiefel manifold as a submanifold of Euclidean space in
\S\ref{sec:eucstiefel}.  This introduces elementary ideas
from differential geometry and provides the geometric structure
of the orthogonal group (a special case of the Stiefel manifold),
which will be used throughout the rest of the paper.  However,
the Euclidean metric is not natural for the Stiefel manifold,
which inherits a canonical metric from its definition as a
quotient space.  Therefore, we introduce the quotient space point
of view in \S\ref{sec:gqs}.  With this viewpoint, we then
derive our formulas for geodesics and parallel translation for
the Stiefel and Grassmann manifold in \S\ref{sec:stiefgeom} and 
\S\ref{sec:grassgeom}.  Finally, we
describe how to incorporate these formulae into conjugate
gradient and Newton methods in \S\ref{sec:cgnmrm}.

\subsection{Manifolds Arising in Numerical Linear Algebra}
\label{sec:def}

For simplicity of exposition, but for no fundamental reason, we
will concentrate on real matrices.  All ideas carry over
naturally to complex matrices. Spaces of interest are

\begin{remunerate}
\item The orthogonal group $O_{n}$ consisting of $n\by n$
  orthogonal matrices
\item The Stiefel manifold $V_{n,p}$ consisting of $n \by p$
  ``tall-skinny'' orthonormal matrices
\item The Grassmann manifold $G_{n,p}$ obtained by identifying
  those matrices in $V_{n,p}$ whose columns span the same
  subspace (a quotient manifold).
\end{remunerate}

Table~\ref{tab:rep} summarizes the definitions of these spaces.
Our description of $G_{n,p}$ is necessarily more abstract than
$O_{n}$ or $V_{n,p}$.  $G_{n,p}$ may be defined as the set of all
$p$-dimensional subspaces of an $n$-dimensional space.

We shall benefit from two different yet equivalent modes of
describing our spaces: concrete representations and quotient
space representations.  Table \ref{tab:comprep} illustrates how
we store elements of $V_{n,p}$ and $G_{n,p}$ in a computer.  A
point in the Stiefel manifold $V_{n,p}$ is represented by an $n
\by p$ matrix.  A point on the Grassmann manifold $G_{n,p}$ is a
linear subspace, which may be specified by an arbitrary
orthogonal basis stored as an $n\by p$ matrix.  An important
difference here is that, unlike points on the Stiefel manifold,
the choice of matrix is not unique for points on the Grassmann
manifold.

\begin{table}
\caption{{\it Representations of subspace manifolds}}
\label{tab:rep}
\boxer{\footnotesize
\begin{tabular}{lccc}
{\sc Space} & {\sc Symbol} & {\sc Matrix Rep.} & {\sc Quotient Rep.} \\
\\ {\bf \large Orthogonal Group} & $ O_{n}$ & $\boxedQ$ & --\\ \\ {\bf
\large Stiefel Manifold} & $V_{n,p}$ & $\boxedY$ & $O_{n}/O_{n-p}$
\\ \\
{\bf \large Grassmann Manifold} & $G_{n,p}$ & None & $\left\{\begin{array}{c}
V_{n,p}/O_{p} \\
\hbox{or} \\ 
O_{n}/\left( O_{p} \times O_{n-p} \right)
\end{array}
\right\} $\\
\end{tabular}
}
\end{table}

The second mode of representation, the more mathematical, is
useful for obtaining closed-form expressions for the geometrical
objects of interest.  It is also the ``proper'' theoretical
setting for these manifolds.  Here, we represent the manifolds as
quotient spaces.  Points in the Grassmann manifold are
equivalence classes of $n \by p$ orthogonal matrices, where two
matrices are equivalent if their columns span the same
$p$-dimensional subspace.  Equivalently, two matrices are
equivalent if they are related by right multiplication of an
orthogonal $p \by p$ matrix.  Therefore,
${G_{n,p}=V_{n,p}/O_{p}}$.  On the computer, by necessity, we
must pick a representative of the equivalence class to specify a
point.

The Stiefel manifold may also be defined as a quotient space but
arising from the orthogonal group.  Here we identify two
orthogonal matrices if their first $p$ columns are identical or,
equivalently, if they are related by right multiplication of a
matrix of the form $\bigl({I\atop0}\;{0\atop Q}\bigr)$, where $Q$
is an orthogonal $(n-p)\by (n-p)$ block.  Therefore,
$V_{n,p}=O_{n}/O_{n-p}$.  With the Stiefel manifold so
represented, one has yet another representation of the Grassmann
manifold, $G_{n,p}=O_{n}/(O_{p}\times O_{n-p})$.

\begin{table}
\caption{{\it Computational representation of subspace manifolds}}
\label{tab:comprep}
\boxer{
\begin{center}\footnotesize
\begin{tabular}{lc@{\hspace*{-.2in}}c@{\hspace*{-.1in}}c}
{\sc Space} & {\sc Data Structure} & {\sc Represents} & {\sc Tangents $\Delta$}
  \\ Stiefel
Manifold & $\boxedY$ & one point & $Y\T\Delta=\hbox{skew-symmetric}$
 \\ \\ Grassmann Manifold & $\boxedY$
& entire equivalence class & $Y\T\Delta=0$ \\
\end{tabular}
\end{center}
}
\end{table}

\subsection{The Stiefel Manifold in Euclidean Space}
\label{sec:eucstiefel}

The Stiefel manifold $V_{n,p}$ may be embedded in the $np$
dimensional Euclidean space of $n \by p$ matrices.  When $p=1$,
we simply have the sphere, while when $p=n$, we have the group of
orthogonal matrices known as $O_{n}$.  These two special cases
are the easiest, and arise in numerical linear algebra the most
often.

Much of this section, which consists of three subsections, is
designed to be a painless and intuitive introduction to
differential geometry in Euclidean space.
\S\ref{sec:tandn} is elementary.  It derives formulas for
projections onto the tangent and normal spaces.  In
\S\ref{sec:geodesics}, we derive formulas for geodesics on
the Stiefel manifold in Euclidean space.  We then discuss
parallel translation in \S\ref{sec:xport}.

In the two special cases when $p=1$ and $p=n$, the Euclidean
metric and the canonical metric to be discussed in
\S\ref{sec:stiefgeom} are the same.  Otherwise they differ.

\subsubsection{Tangent and Normal Space}
\label{sec:tandn}

\begin{figure}[htb]
\centerline{\epsfbox[23 612 350 777]{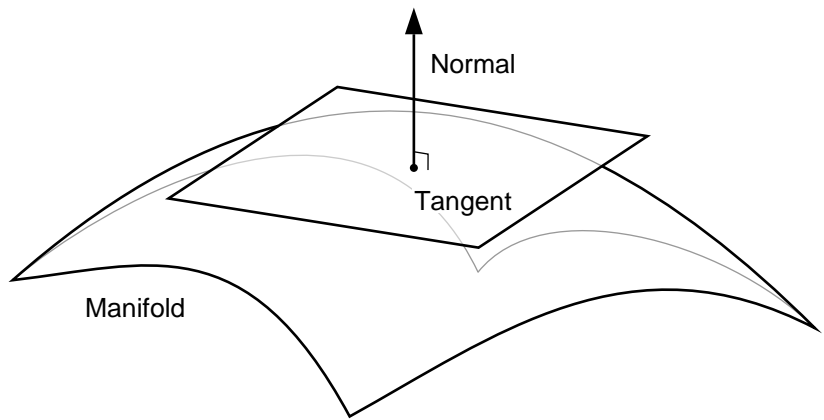}}
\caption{The tangent and normal spaces of an embedded or constraint manifold.}
\label{fig:TN}
\end{figure}

Intuitively, the tangent space at a point is the plane tangent to
the submanifold at that point, as shown in Figure~\ref{fig:TN}.
For $d$-dimensional manifolds, this plane is a $d$-dimensional
vector space with origin at the point of tangency.  The normal
space is the orthogonal complement.  On the sphere, tangents are
perpendicular to radii, and the normal space is radial.  In this
subsection, we will derive the equations for the tangent and
normal spaces on the Stiefel manifold.  We also compute the
projection operators onto these spaces.

An equation defining tangents to the Stiefel manifold at a point
$Y$ is easily obtained by differentiating $Y^TY=I$, yielding
$Y^T\!\Delta +\Delta^T Y=0$, i.e., $Y\T\Delta$ is skew-symmetric.
This condition imposes $p(p+1)/2$ constraints on $\Delta$, or
equivalently, the vector space of all tangent vectors $\Delta$
has dimension
\begin{equation} np-{p(p+1)\over2}={p(p-1)\over2}+p(n-p).
\label{eq:dofV} \end{equation} Both sides of Eq.~(\ref{eq:dofV}) are
useful for the dimension counting arguments that will be employed.

The normal space is defined to be the orthogonal complement of
the tangent space.  Orthogonality depends upon the definition of
an inner product, and because in this sub-section we view the
Stiefel manifold as an embedded manifold in Euclidean space, we
choose the standard inner product \begin{equation}
  g_e(\Delta_1,\Delta_2) = \tr\Delta_1^T\!\Delta_2,
  \label{eq:Euclmetric}\end{equation} in $np$-dimensional
Euclidean space (hence the subscript $e$), which is also the
Frobenius inner product for $n\by p$ matrices.  We shall also
write $\langle\Delta_1,\Delta_2\rangle$
for the inner product, which may or may not be the Euclidean one.
The normal space at a point $Y$ consists of all matrices $N$
which satisfy
$$\tr\Delta^T N=0$$ for all $\Delta$ in the tangent space.  It follows
that the normal space is $p(p+1)/2$ dimensional.  It is easily
verified that if $N=YS$, where $S$ is $p\by p$ symmetric, then $N$ is
in the normal space.  Since the dimension of the space of such
matrices is $p(p+1)/2$, we see that the normal space is exactly the
set of matrices $\{\,YS\,\}$, where $S$ is any $p \by p$ symmetric
matrix.

Let $Z$ be any $n\by p$ matrix.  Letting $\sym(A)$ denote
$(A+A^T)/2$ and $\skew(A)=(A-A^T)/2$, it is easily verified that
at $Y$
\begin{equation}\pi_N(Z)= Y\sym(Y^T\!Z)
\label{eq:Np}\end{equation} defines a
projection of $Z$ onto the normal space.  Similarly, at $Y$,
\begin{equation}\pi_T(Z)= Y \skew(Y^T\!Z) +({I-YY^T})Z \label{eq:piT}
\end{equation} is a projection of $Z$ onto the tangent space
at~$Y$ (this is also true of the canonical metric to be discussed in
\S\ref{sec:stiefgeom}). Eq.~(\ref{eq:piT}) suggests a form for
the tangent space of~$V_{n,p}$ at~$Y$ that will prove to be
particularly useful.  Tangent directions $\Delta$ at~$Y$ then have the
general form, \begin{eqnarray} \Delta&=&YA+\Yperp B \label{eq:TV} \\
&=& YA+(I-YY^T)C \label{eq:TV2}\end{eqnarray} where $A$ is $p\by p$
skew-symmetric, $B$ is $(n-p)\by p$, $C$ is $n\by p$, $B$ and $C$ are
both arbitrary, and $\Yperp$ is any $n\by(n-p)$ matrix such that
$YY^T+\Yperp\Yperp^T=I$; note that $B=\Yperp^TC$.  The entries in the
matrices $A$ and $B$ parameterize the tangent space at~$Y$ with
$p(p-1)/2$ degrees of freedom in~$A$ and $p(n-p)$ degrees of freedom
in~$B$, resulting in $p(p-1)/2+p(n-p)$ degrees of freedom as seen in
Eq.~(\ref{eq:dofV}).

In the special case $Y=I_{n,p}\equiv{I_p\atopwithdelims()0}$ (the
first $p$ columns of the $n\by n$ identity matrix), called the
origin, the tangent space at~$Y$ consists of those matrices $$X =
{A\atopwithdelims()B},$$ for which $A$ is $p\by p$ skew-symmetric
and $B$ is $(n-p)\by p$ arbitrary.

\subsubsection{Embedded Geodesics}
\label{sec:geodesics}

A geodesic is the curve of shortest length between two points on
a manifold.  A straightforward exercise from the calculus of
variations reveals that {\em for the case of manifolds embedded
  in Euclidean space\/} the acceleration vector at each point
along a geodesic is normal to the submanifold so long as the
curve is traced with uniform speed.  This condition is necessary
and sufficient.  In the case of the sphere, acceleration for
uniform motion on a great circle is directed radially and
therefore normal to the surface; therefore, great circles are
geodesics on the sphere.  One may consider embedding manifolds in
spaces with arbitrary metrics.  See Spivak \cite[vol.~3,
p.~4]{spivak79} for the appropriate generalization.

Through Eq.~(\ref{eq:Np}) for the normal space to the Stiefel
manifold, it is easily shown that the geodesic equation for a
curve $Y(t)$ on the Stiefel manifold is defined by the
differential equation
\begin{equation} \label{eq:motion} \Ydd+Y(\Yd^T\Yd)=0.
\end{equation} To see this, we begin
with the condition that $Y(t)$ remains on the Stiefel manifold, 
\begin{equation}
\label{eq:g0} Y^TY=I_p. \end{equation} Taking two derivatives, 
\begin{equation} \label{eq:g1}
  Y^T\Ydd+2\Yd^T\Yd+\Ydd^TY=0. \end{equation} To be a geodesic,
$\Ydd(t)$ must be in the normal space at $Y(t)$ so that
\begin{equation} \label{eq:g2} \Ydd(t)+Y(t)S=0 \end{equation} for
some symmetric matrix $S$.  Substitute Eq.~(\ref{eq:g2}) into
(\ref{eq:g1}) to obtain the geodesic equation
Eq.~(\ref{eq:motion}).  Alternatively Eq.~(\ref{eq:motion}) could
be obtained from the Euler-Lagrange equation for the calculus of
variations problem \begin{equation} d(Y_1,Y_2)
  =\min_{Y(t)}\int_{t_1}^{t_2}(\tr\Yd^T\Yd)^{1/2}\,dt\quad
  \hbox{such that $Y(t_1)=Y_1$,
    $Y(t_2)=Y_2$.}\label{eq:E-L}\end{equation}

We identify three integrals of motion of the geodesic equation
Eq.~(\ref{eq:motion}).  Define \begin{equation}
  C=Y^TY,\qquad A=Y^T\Yd,\qquad S=\Yd^T\Yd. \end{equation}
Directly from the geodesic equation Eq.~(\ref{eq:motion}),
\begin{eqnarray*} \dot{C}&=&A+A^T,\\ 
  \dot{A}&=&-CS+S,\\ \dot{S}&=&[A,S], \end{eqnarray*} where
\begin{equation}[A,S]=AS-SA\end{equation} is the Lie bracket of
two matrices.  Under the initial conditions that $Y$ is on the
Stiefel manifold ($C=I$) and $\Yd$ is a tangent ($A$ is
skew-symmetric) then the integrals of the motion have the values
\begin{eqnarray*} C(t)&=&I,\\ A(t)&=&A(0),\\ 
  S(t)&=&e^{At}S(0)e^{-At}. \end{eqnarray*} These integrals
therefore identify a constant speed curve on the Stiefel
manifold.  In most differential geometry books, the equation of
motion for geodesics is written in intrinsic coordinates in terms
of so-called Christoffel symbols which specify a quadratic form
of the tangent vectors.  In our formulation, the form
$\Gamma_e(\Yd,\Yd)=Y\Yd^T \Yd$ is written compactly in extrinsic
coordinates.

With these constants of the motion, we can write an integrable
equation for the final geodesic,\footnote{We thank Ross Lippert
  \cite{lippert95b} for this observation.} $$\frac{d}{dt}
\left(Ye^{At} \matsep \Yd e^{At}\right)=\left(Ye^{At} \matsep \Yd
  e^{At}\right) \pmatrix{A&-S(0)\cr I&A\cr},$$ with integral
$$Y(t)=\left(Y(0) \matsep \Yd(0) \right)\; \exp t
\pmatrix{A&-S(0)\cr I&A\cr}\; I_{2p,p} e^{-At}.$$

This is an exact closed form expression for the geodesic on the
Stiefel manifold, but we will not use this expression in our
computation.  Instead we will consider the non-Euclidean
canonical metric on the Stiefel manifold in
\S\ref{sec:stiefgeom}.

We mention in the case of the orthogonal group ($p=n$), the
geodesic equation is obtained simply from
$A=Q^T\!\dot{Q}={}$constant, yielding the simple solution
\begin{equation} Q(t)=Q(0) e^{At}.\label{eq:SOgeo}\end{equation}
From Eq.~(\ref{eq:SOgeo}) it is straightforward to show that on
connected components of~$O_n$, \begin{equation} d(Q_1,Q_2)
  =\biggl(\sum_{k=1}^n\theta_k^2\biggr)^{1/2}\end{equation} where
$\{e^{i\theta_k}\}$ are the eigenvalues of the matrix $Q_1^TQ_2$
(cf.\ Eq.~(\ref{eq:arclengthd}) and \S\ref{sec:csdecomp}).

\subsubsection{Parallel Translation}
\label{sec:xport}

In Euclidean space, we move vectors parallel to themselves simply by
moving the base of the arrow.  On an embedded manifold, if we move a
tangent vector to another point on the manifold by this technique, it
is generally not a tangent vector.  One can, however, transport
tangents along paths on the manifold by infinitesimally removing the
component of the transported vector in the normal space.

\begin{figure}[htb]
\centerline{\epsfbox[11 616 297 757]{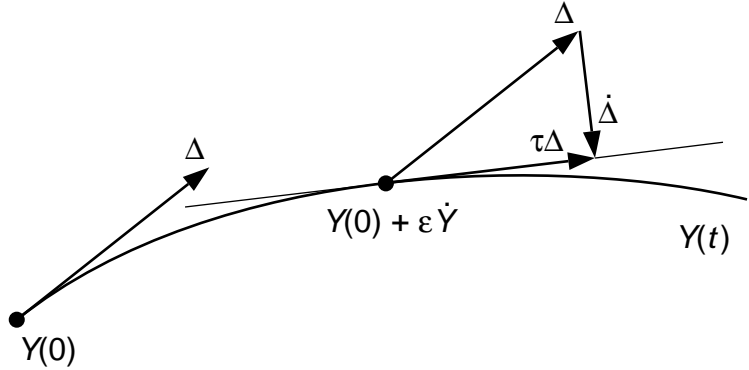}}
\caption{Parallel transport in a submanifold  of Euclidean space 
(infinitesimal construction).}
\label{fig:Tp}
\end{figure}

Figure~\ref{fig:Tp} illustrates the idea: Imagine moving a
tangent vector $\Delta$ along the curve $Y(t)$ in such a manner
that every infinitesimal step consists of a parallel displacement
of $\Delta$ in the Euclidean $np$ dimensional space which is then
followed by the removal of the normal component.  If we move from
$Y(0)= Y $ to $Y(\epsilon)$ then to first order, our new location
is $Y+ \epsilon \Yd$.  The equation for infinitesimally removing
the component generated in the normal space as we move in the
direction $\Yd$ is obtained by differentiating Eq.~(\ref{eq:Np}):
\begin{equation}\dot\Delta =-Y(\Yd^T\!\Delta
  +\Delta^T\Yd)/2,\label{eq:ptstiefeuc}\end{equation} We are
unaware of any closed form solution to this system of
differential equations along geodesics.

By differentiation, we see that parallel transported vectors preserve
the inner product.  In particular, the square length of $\Delta$ ($\tr
\Delta^T\!\Delta$) is preserved.  Additionally, inserting $\Yd$ into
the parallel transport equation, one quickly sees that a geodesic
always parallel transports its own tangent vector.  This condition may
be taken as the definition of a geodesic.

Observing that $\tr \Delta^T\!\Delta$ is the sum of the squares of the
singular values of $\Delta$, we conjectured that the individual
singular values of $\Delta$ might also be preserved by parallel
transport.  Numerical experiments show that this is not the case.

In the case of the orthogonal group ($p=n$), however, parallel
translation of $\Delta$ along the geodesic $Q(t)=Q(0) e^{At}$ is
straightforward.  Let $\Delta(t)=Q(t)B(t)$ be the solution of the
parallel translation equation $$\dot\Delta=-Q(\Qd^T\!\Delta
+\Delta^T\Qd)/2,$$ where $B(t)$ is a skew-symmetric matrix.
Substituting $\dot{\Delta}=\Qd B+Q \dot{B}$ and $\Qd=QA$, we
obtain
\begin{equation} \dot{B}=-\frac{1}{2}[A,B] \label{eq:PTOn}\end{equation}
whose solution is $B(t)=e^{-At/2}B(0)e^{At/2}$; therefore,
\begin{equation}\Delta(t)=Q(0)e^{At/2}B(0)e^{At/2}.
\end{equation}  These formulas may be
generalized to arbitrary connected Lie groups \cite[Chap.~2,
Ex.~A.6]{helgason78}.

So as to arrive at the general notion of parallel transport, let us
formalize what we did here.  We saw that the geodesic equation may be
written $$\Ydd+\Gamma_e(\Yd,\Yd)=0,$$ where in the Euclidean case
$$\Gamma_e(\Delta_1,\Delta_2)= Y(\Delta_1^T \Delta_2 + \Delta_2^T
\Delta_1)/2.$$ Anticipating the generalization, we interpret $\Gamma$
as containing the information of the normal component that needs to be
removed.  Knowing the quadratic function $\Gamma(\Delta,\Delta)$ is
sufficient for obtaining the bilinear function
$\Gamma(\Delta_1,\Delta_2)$; the process is called polarization.  We
assume that $\Gamma$ is a symmetric function of its arguments (this is
the so-called torsion-free condition), and we obtain
$$4\Gamma(\Delta_1,\Delta_2)=\Gamma(\Delta_1+\Delta_2,
\Delta_1+\Delta_2)-\Gamma(\Delta_1-\Delta_2,\Delta_1-\Delta_2).$$ For
the cases we study in this paper, it is easy in practice to guess a
symmetric form for $\Gamma(\Delta_1,\Delta_2)$ given
$\Gamma(\Delta,\Delta)$.

We will give a specific example of why this formalism is needed in
\S\ref{sec:stiefgeom}. Let us mention here that the parallel
transport defined in this manner is known to differential geometers as
the Levi-Civita connection.  We also remark that the function $\Gamma$
when written in terms of components defines the Christoffel symbols.
Switching to vector notation, in differential geometry texts the $i$th
component of the function $\Gamma(v,w)$ would normally be written as
$\sum_{jk} \Gamma_{jk}^i v_j w_k$, where the constants $\Gamma_{jk}^i$
are called Christoffel symbols.  We prefer the matrix notation over
the scalar notation.

\subsection{Geometry of Quotient Spaces}
\label{sec:gqs}

Given a manifold whose geometry is well understood (where there are
closed form expressions for the geodesics and, perhaps also, parallel
transport), there is a very natural, efficient, and convenient way to
generate closed form formulas on quotient spaces of that manifold.
This is precisely the situation with the Stiefel and Grassmann
manifolds, which are quotient spaces within the orthogonal group.  As
just seen in the previous section, geodesics and parallel translation
on the orthogonal group are simple.  We now show how the Stiefel and
Grassmann manifolds inherit this simple geometry.

\subsubsection{The Quotient Geometry of the Stiefel Manifold}
\label{sec:qgstief}

The important ideas here are the notions of the horizontal and
vertical spaces, the metric, and their relationship to geodesics
and parallel translation.  We use brackets to denote equivalence
classes.  We will define these concepts using the Stiefel
manifold $V_{n,p}=O_{n}/O_{n-p}$ as an example. The equivalence
class $[Q]$ is the set of all $n\by n$ orthogonal matrices with
the same first $p$ columns as~$Q$.  A point in the Stiefel
manifold is the equivalence class \begin{equation} [Q]=\left\{\,
    Q\pmatrix{I_p&0\cr0&Q_{n-p}\cr} : Q_{n-p}\in
    O_{n-p}\,\right\},\label{eq:stiefeleq}\end{equation} that is,
a point in the Stiefel manifold is a particular subset of the
orthogonal matrices.  Notice that in this section we are working
with equivalence classes rather than $n\by p$ matrices
$Y=QI_{n,p}$.

The vertical and horizontal spaces at a point $Q$ are complementary
linear subspaces of the tangent space at~$Q$.  The vertical space is
defined to be vectors tangent to the set $[Q]$.  The horizontal space
is defined as the tangent vectors at~$Q$ orthogonal to the vertical
space.  At a point $Q$, the vertical space is the set of vectors of
the form \begin{equation}\Phi =
  Q\pmatrix{0&0\cr0&C\cr},\label{eq:stiefV}
\end{equation} where
$C$ is $(n-p)\by(n-p)$ skew-symmetric, and we have hidden
post-multiplication by the isotropy subgroup
$\left({\smash{I_p}\atop}\; {\atop \smash{O_{n-p}}}\right)$.
Such vectors are clearly tangent to the set $[Q]$ defined in
Eq.~(\ref{eq:stiefeleq}).  It follows that the horizontal space
at $Q$ is the set of tangents of the form \begin{equation}\Delta
  = Q\pmatrix{A&-B^T\cr B&0\cr}\label{eq:stiefH}\end{equation}
(also hiding the isotropy subgroup), where $A$ is $p \by p$
skew-symmetric.  Vectors of this form are clearly orthogonal to
vertical vectors with respect to the Euclidean inner product.
The matrices $A$ and $B$ of Eq.~(\ref{eq:stiefH}) are equivalent
to those of Eq.~(\ref{eq:TV}).

The significance of the horizontal space is that it provides a
representation of tangents to the quotient space.  Intuitively,
movements in the vertical direction make no change in the quotient
space.  Therefore, the metric, geodesics, and parallel translation
must all be restricted to the horizontal space.  A rigorous treatment
of these intuitive concepts is given by Kobayashi and
Nomizu~\cite{kobayashi} and Chavel~\cite{chavel93}.

The canonical metric on the Stiefel manifold is then simply the
restriction of the orthogonal group metric to the horizontal space
(multiplied by $1/2$ to avoid factors of~$2$ later on).  That is, for
$\Delta_1$ and $\Delta_2$ of the form in Eq.~(\ref{eq:stiefH}), 
\begin{eqnarray}
g_c(\Delta_1,\Delta_2)&=&{1\over2} \tr\left(Q\pmatrix{A_1&-B_1^T\cr
B_1&0\cr}\right)^T Q\pmatrix{A_2&-B_2^T\cr B_2&0\cr}\\ &=& \half\tr
A_1^T\!A_2 +\tr B_1^T\!B_2,\nonumber
\label{eq:stiefcanonm}
\end{eqnarray} which we shall
also write as $\langle\Delta_1,\Delta_2\rangle$.  It is important to
realize that this is {\em not\/} equal to the Euclidean metric $g_e$
defined in \S\ref{sec:eucstiefel} (except for $p=1$ or $n$),
even though we use the Euclidean metric for the orthogonal group in
its definition.  The difference arises because the Euclidean metric
counts the independent coordinates of the skew-symmetric $A$ matrix
twice and those of~$B$ only once, whereas the canonical metric counts
all independent coordinates in $A$ and $B$ equally.  This point is
discussed in detail in \S\ref{sec:stiefgeom}.

Notice that the orthogonal group geodesic
\begin{equation} \label{gof} Q(t)=Q(0) \exp t \pmatrix{A&-B^T\cr B&0\cr}
\label{eq:orthogeodstief}\end{equation} has horizontal tangent
\begin{equation}\dot Q(t)=Q(t)\pmatrix{A&-B^T\cr B&0\cr}
\end{equation} at every point along
the curve $Q(t)$.  Therefore they are curves of shortest length in the
quotient space as well, i.e., geodesics in the Grassmann manifold are
given by the simple formula \begin{equation} 
\hbox{Stiefel geodesics} =[Q(t)],\end{equation}
where $[Q(t)]$ is given by Eqs.\ (\ref{eq:stiefeleq}) and
(\ref{eq:orthogeodstief}).  This formula will be essential for
deriving an expression for geodesics on the Stiefel manifold using
$n\by p$ matrices in \S{}\ref{sec:stiefgeom}.

In a quotient space, parallel translation works in a way similar to
the embedded parallel translation discussed in
\S{}\ref{sec:xport}.  Parallel translation along a curve (with
everywhere horizontal tangent) is accomplished by infinitesimally
removing the vertical component of the tangent vector.  The equation
for parallel translation along the geodesics in the Stiefel manifold
is obtained by applying this idea to Eq.~(\ref{eq:PTOn}), which
provides translation along geodesics for the orthogonal group.  Let
\begin{equation} \Afrak =\pmatrix{A_1&-B_1^T\cr B_1&0\cr}\quad\hbox{and}\quad
\Bfrak=\pmatrix{A_2&-B_2^T\cr B_2&0\cr}\end{equation} be two horizontal vectors
at~$Q=I$.  The parallel translation of $\Bfrak$ along the geodesic
$e^{\Afrak t}$ is given by the differential equation \begin{equation} \dot\Bfrak =
-\frac{1}{2}[\Afrak,\Bfrak]_H, \label{eq:rhsptg}\end{equation} where the
subscript $H$ denotes the horizontal component (lower right block set
to zero).  Note that the Lie bracket of two horizontal vectors is not
horizontal, and that the solution to Eq.~(\ref{eq:rhsptg}) is not
given by the formula $(e^{-\Afrak t/2}\Bfrak(0)e^{\Afrak t/2})_H$.
This is a special case of the general formula for reductive
homogeneous spaces \cite{chavel93,smith93a}.  This first order linear
differential equation with constant coefficients is integrable in
closed form, but it is an open question whether this can be accomplished
with $O(np^2)$ operations.

\subsubsection{The Quotient Geometry of the Grassmann Manifold}
\label{sec:qggrass}

We quickly repeat this approach for the Grassmann manifold
$G_{n,p}=O_{n}/(O_{p}\times O_{n-p})$.  The equivalence class $[Q]$ is
the set of all orthogonal matrices whose first $p$ columns span the
same subspace as those of~$Q$.  A point in the Grassmann manifold is
the equivalence class \begin{equation}[Q]=\left\{\, Q\pmatrix{Q_p&0\cr0&Q_{n-p}\cr} :
Q_p \in O_p, \; Q_{n-p} \in O_{n-p} \,\right\},
\label{eq:grasseqclass}\end{equation} i.e., a point in the Grassmann manifold is
a particular subset of the orthogonal matrices, and the Grassmann
manifold itself is the collection of all these subsets.

The vertical space at a point $Q$ is the set of vectors of the form
\begin{equation}\Phi = Q\pmatrix{A&0\cr 0&C\cr},\label{eq:grassV}\end{equation} where $A$ is
$p\by p$ skew-symmetric and $C$ is $(n-p)\by(n-p)$ skew-symmetric.
The horizontal space at $Q$ is the set of matrices of the form
\begin{equation}\Delta = Q\pmatrix{0&-B^T\cr B&0\cr}.\label{eq:grassH}\end{equation} Note that
we have hidden post-multiplication by the isotropy subgroup
$\left({\smash{O_p}\atop}\;{\atop \smash{O_{n-p}}}\right)$ in Eqs.\
(\ref{eq:grassV}) and (\ref{eq:grassH}).

The canonical metric on the Grassmann manifold is the restriction of
the orthogonal group metric to the horizontal space (multiplied
by~$1/2$).  Let $\Delta_1$ and $\Delta_2$ be of the form in
Eq.~(\ref{eq:grassH}).  Then \begin{equation} g_c(\Delta_1,\Delta_2) =\tr
B_1^T\!B_2.\end{equation} As opposed to the canonical metric for the Stiefel
manifold, this metric is in fact equivalent to the Euclidean metric
(up to multiplication by $1/2$) defined in Eq.~(\ref{eq:Euclmetric}).

The orthogonal group geodesic
\begin{equation} \label{gofgrassmann} Q(t)=Q(0) \exp t\pmatrix{0&-B^T\cr B&0\cr}
\label{eq:grassgeod}\end{equation} has horizontal tangent
\begin{equation}\dot Q(t)=Q(t)\pmatrix{0&-B^T\cr B&0\cr}\end{equation} at every point along
the curve $Q(t)$; therefore, \begin{equation}\hbox{Grassmann geodesics} =
[Q(t)],\label{eq:grassqgeod}\end{equation} where $[Q(t)]$ is given by
Eqs.\ (\ref{eq:grasseqclass}) and (\ref{eq:grassgeod}).  This formula
gives us an easy method for computing geodesics on the Grassmann
manifold using $n\by p$ matrices, as will be seen in
\S{}\ref{sec:grassgeom}.

The method for parallel translation along geodesics in the Grassmann
manifold is the same as for the Stiefel manifold, although it turns
out the Grassmann manifold has additional structure that makes this
task easier.  Let \begin{equation} \Afrak =\pmatrix{0&-A^T\cr A&0\cr}
\quad\hbox{and}\quad \Bfrak=\pmatrix{0&-B^T\cr B&0\cr}\end{equation} be two
horizontal vectors at~$Q=I$.  It is easily verified that
$[\Afrak,\Bfrak]$ is in fact a vertical vector of the form of
Eq.~(\ref{eq:grassV}).  If the vertical component of
Eq.~(\ref{eq:PTOn}) is infinitesimally removed, we are left with the
trivial differential equation \begin{equation} \dot\Bfrak = 0.
\label{eq:grassptde}\end{equation} Therefore, the parallel translation of the
tangent vector $Q(0)\Bfrak$ along the geodesic $Q(t)=Q(0)e^{\Afrak t}$
is simply given by the expression \begin{equation} \tau\Bfrak(t) =Q(0)e^{\Afrak
t}\Bfrak, \label{eq:ptgrassq}\end{equation} which is of course horizontal
at~$Q(t)$.  Here we introduce the notation $\tau$ to indicate the
transport of a vector; it is not a scalar multiple of the vector.  It
will be seen in \S{}\ref{sec:grassgeom} how this formula may be
computed using $O(np^2)$ operations.

As an aside, if $H$ and $V$ represent the horizontal and vertical
spaces, respectively, it may be verified that
\begin{equation} [V,V]\subset V,\quad [V,H]\subset H,\quad
  [H,H]\subset V.
\end{equation} The
first relationship follows from the fact that $V$ is a Lie algebra,
the second follows from the reductive homogeneous space structure
\cite{kobayashi} of the Grassmann manifold, also possessed by the
Stiefel manifold, and the third follows the symmetric space structure
\cite{helgason78,kobayashi} of the Grassmann manifold, which
the Stiefel manifold does not possess.

\goodbreak
\samepage{
\subsection{The Stiefel Manifold with its Canonical Metric}
\label{sec:stiefgeom}

\subsubsection{The Canonical Metric (Stiefel)}

The Euclidean metric $$g_e(\Delta,\Delta) =\tr\Delta^T\!\Delta$$ used
in \S{}\ref{sec:eucstiefel} may seem natural, but one reasonable
objection to its use is that it weighs the independent degrees of
freedom of the tangent vector unequally.  Using the representation of
tangent vectors $\Delta=YA+\Yperp B$ given in Eq.~(\ref{eq:TV}), it is
seen that \begin{eqnarray*} g_e (\Delta,\Delta)&= &\tr A^T\!A +\tr
B^T\!B.\\ &=&2\sum_{i<j} a_{ij}^2 +\sum_{ij} b_{ij}^2 \end{eqnarray*}
The Euclidean metric counts the $p(p+1)/2$ independent coordinates of
$A$ twice.  At the origin $I_{n,p}$, a more equitable metric would be
$g_c(\Delta,\Delta) =\tr\Delta^T(I-\half I_{n,p}I_{n,p}^T)\Delta =
\half\tr A^T\!A + \tr B^T\!B$.  To be equitable at all points in the manifold,
the metric must vary with $Y$ according to \begin{equation} \label{canon}
g_c(\Delta,\Delta) =\tr\Delta^T(I-\half YY^T)\Delta. \label{eq:stiefgc}\end{equation}
This is called the {\em canonical metric\/} on the Stiefel manifold.
This is precisely the metric derived from the quotient space structure
of $V_{n,p}$ in Eq.~(\ref{eq:stiefcanonm}); therefore, the formulas
for geodesics and parallel translation for the Stiefel manifold given
in \S{}\ref{sec:qgstief} are correct if we view the Stiefel
manifold as the set of orthonormal $n\by p$ matrices with the metric of
Eq.~(\ref{eq:stiefgc}).  Note that if $\Delta=YA+\Yperp B$ is a
tangent vector, then $g_c(\Delta,\Delta)=\half\tr A^T\!A +\tr B^T\!B$, as
seen previously.
}

\subsubsection{Geodesics (Stiefel)}

The path length \begin{equation} L=\int g_c(\Yd,\Yd)^{1/2}\,dt
\label{eq:pathlength}\end{equation} may be minimized with the calculus of
variations.  Doing so is tedious but yields the new geodesic equation
\begin{equation} \Ydd +\Yd\Yd^TY +Y\bigl((Y^T\Yd)^2+\Yd^T\Yd\bigr) =0,
\label{eq:cgeod} \end{equation} Direct substitution into
Eq.~(\ref{eq:cgeod}) using the fact that
$$(I-I_{n,p}I_{n,p}^T)X(I-I_{n,p}I_{n,p}^T)=0,$$ if $X$ is a
skew-symmetric matrix of the form $$X=\pmatrix{A&-B^T\cr B&0\cr},$$
verifies that the paths of the form \begin{equation} Y(t)=Q e^{Xt} I_{n,p}
\label{cgo} \end{equation} are closed form solutions to the
geodesic equation for the canonical metric.

We now turn to the problem of computing geodesics with algorithms of
complexity $O(np^2)$.  Our current formula $Y(t) =
Q\exp{t\left({A\atop B}\;{-{B^T}\atop0}\right)} I_{n,p}$ for a
geodesic is not useful.  Rather we want to express the geodesic $Y(t)$
in terms of the current position $Y(0)=Y$ and a direction $\Yd(0)=H$.
For example, $A=Y\T H$ and we have $C:=B\T B= H\T (I-YY^T)H$.  In fact
the geodesic only depends on $B\T B$ rather than $B$ itself.  The
trick is to find a differential equation for $M(t)=
I_{n,p}^T\exp{t\left({A\atop B}\;{-{B^T}\atop0}\right)} I_{n,p}.$

The following theorem makes clear that the computational difficulty
inherent in computing the geodesic is the solution of a constant
coefficient second order differential equation for $M(t)$.  The answer
is obtained not by a differential equation solver but rather by solving
the corresponding quadratic eigenvalue problem:

\begin{theorem}
If $Y(t)=Qe^{t\left({A\atop B}\;{-\smash{B^T}\atop0}\right)} I_{n,p}$,
with $Y(0)=Y$ and $\Yd(0)=H$, then \begin{equation} Y(t)=Y M(t) +(I-YY^T)H
\int_0^t M(t)\,dt, \label{eq:stiefgeo} \end{equation} where $M(t)$ is the
solution to the second order differential equation with constant
coefficients, \begin{equation} \label{odem} {\ddot{M}}-A\dot{M} +CM =0;\qquad
M(0)=I_p,\quad \dot M(0)=A, \label{eq:M(t)de} \end{equation} $A=Y\T H $ is
skew-symmetric, and $C=H\T (I-YY^T)H$ is non-negative definite.
\label{th:stiefgeo}
\end{theorem}

{\em Proof}. A direct computation verifies that
$M(t)$ satisfies Eq.~(\ref{odem}). By separately considering $Y^T
Y(t)$ and $(I-YY^T)Y(t)$, we may derive Eq.~(\ref{eq:stiefgeo}).
\qquad\endproof


The solution of the differential equation Eq.~(\ref{odem}) may be
obtained \cite{datta95,wilk65} by solving the quadratic eigenvalue
problem $$(\lambda^2 I - A \lambda +C)x=0.$$ Such problems are
typically solved in one of three ways: (1)~by solving the generalized
eigenvalue problem $$\pmatrix{C&0\cr0&I\cr}\pmatrix{x\cr\lambda x\cr}
=\lambda\pmatrix{A&-I\cr I&0\cr}\pmatrix{x\cr\lambda x\cr},$$
(2)~by solving the eigenvalue problem $$\pmatrix{0&I\cr-C&A\cr}
\pmatrix{x\cr\lambda x\cr} =\lambda\pmatrix{x\cr\lambda x\cr},$$
or (3)~any equivalent problem obtained by factoring $C=K\T K$ and then
solving the eigenvalue problem $$\pmatrix{A&-K^T\cr
K&0\cr}\pmatrix{x\cr y\cr} =\lambda\pmatrix{x\cr y\cr}.$$

Problems of this form arise frequently in mechanics, usually with $A$
symmetric.  Some discussion of physical interpretations for
skew-symmetric matrices may be found in the context of rotating
machinery \cite{crandall90a}.  If $X$ is the $p \by 2p$ matrix of
eigenvectors and $\Lambda$ denotes the eigenvalues, then
$M(t)=Xe^{\Lambda t}Z$, and its integral is $\int M(t)\,dt=Xe^{\Lambda
t}\Lambda^{-1}Z$, where $Z$ is chosen so that $XZ=I$ and $X\Lambda
Z=A$.

Alternatively, the third method along with the matrix exponential may
be employed:

\begin{corollary} Let $Y$ and $H$ be $n\by p$ matrices such that
$Y^TY=I_p$ and $A=Y\T H$ is skew-symmetric.  Then the geodesic on the
Stiefel manifold emanating from~$Y$ in direction $H$ is given by the
curve \begin{equation} Y(t) =YM(t) +QN(t), \label{eq:sgcomp} 
\end{equation} where \begin{equation}
QR:=K=(I-YY^T)H\end{equation} is the compact QR-decomposition of~$K$ ($Q$ $n\by
p$, $R$ $p\by p$), and $M(t)$ and $N(t)$ are $p\by p$ matrices given
by the matrix exponential \begin{equation} \pmatrix{M(t)\cr N(t)\cr}=\exp
t\pmatrix{A&-R^T\cr R&0\cr} \pmatrix{I_p\cr
  0\cr}. \label{eq:2ksg} 
\end{equation}
\label{cor:sgcomp}
\end{corollary}

\makeatletter
\def\diffeqalign#1{\null\,\vcenter{\openup\jot\m@th
  \ialign{\strut\hfil$\displaystyle{##}$&$\displaystyle{{}##}$\hfil
      &$\displaystyle\qquad##$\hfil\crcr#1\crcr}}\,}
\makeatother


Note that Eq.~(\ref{eq:2ksg}) is easily computed by solving a
$2p\by2p$ skew-symmetric eigenvalue problem, which can be accomplished
efficiently using the SVD or algorithms specially tailored for this
problem \cite{ward78}.

\subsubsection{Parallel Translation (Stiefel)}

We now develop a notion of parallel transport that is consistent
with the canonical metric.  The geodesic equation takes the form
$\Ydd+\Gamma(\Yd,\Yd)=0$, where, from Eq.~(\ref{eq:cgeod}), it is
seen that the Christoffel function for the canonical metric is
\begin{equation} \label{cgc}\Gamma_c(\Delta,\Delta)=\Delta\Delta^T Y + Y
\Delta^T(I-YY^T)\Delta.\end{equation}  By polarizing we obtain the result
\begin{eqnarray}\Gamma_c(\Delta_1,\Delta_2)&=& \half(\Delta_1\Delta_2^T
+\Delta_2\Delta_1^T)Y+\half Y\bigl(\Delta_2\T (I-YY^T)\Delta_1 \\
&& \quad + \Delta_1^T (I-YY^T)\Delta_2\bigr).\nonumber
\label{polo}\end{eqnarray} Parallel transport
is given by the differential equation \begin{equation} \dot\Delta
+\Gamma_c(\Delta,\Yd)=0, \label{eq:stiefpt}\end{equation} which is
equivalent to Eq.~(\ref{eq:rhsptg}).  As stated after this equation,
we do not have an $O(np^2)$ method to compute $\Delta(t)$.

\subsubsection{The Gradient of a Function (Stiefel)}
\label{sec:Vgrad}

Both conjugate gradient and Newton's method require a computation
of the gradient of a function, which depends upon the choice of
metric.  For a function~$F(Y)$ defined on the Stiefel manifold,
the gradient of~$F$ at~$Y$ is defined to be the tangent vector
$\nabla F$ such that
\begin{equation}\tr F_Y\T\Delta =g_c(\nabla F,\Delta) 
\equiv\tr(\nabla F)^T(I-\half
YY^T)\Delta \label{eq:stiefgraddef}\end{equation} for~all tangent vectors
$\Delta$ at~$Y$, where $F_Y$ is the $n\by p$ matrix of partial
derivatives of~$F$ with respect to the elements of~$Y$, i.e.,
\begin{equation} (F_Y)_{ij}={\partial F\over \partial
Y_{ij}}. \label{eq:FYdef} \end{equation} Solving
Eq.~(\ref{eq:stiefgraddef}) for $\nabla F$ such that $Y^T(\nabla
F)={}$skew-symmetric yields \begin{equation} \nabla F =
F_Y-YF_Y^TY. \label{eq:Vnpgradf} \end{equation}
Eq.~(\ref{eq:Vnpgradf}) may also be derived by differentiating
$F\bigl(Y(t)\bigr)$, where $Y(t)$ is the Stiefel geodesic given by
Eq.~(\ref{eq:sgcomp}).

\subsubsection{The Hessian of a Function (Stiefel)}
\label{sec:VHess}

Newton's method requires the Hessian of a function, which depends
upon the choice of metric.  The Hessian of a function $F(Y)$
defined on the Stiefel manifold is defined as the quadratic form
\begin{equation} \Hess F(\Delta,\Delta) =\left.{d^2\over
      dt^2}\right|_{t=0} F\bigl(Y(t)\bigr),
  \label{eq:Hessgdef}\end{equation} where $Y(t)$ is a geodesic
with tangent $\Delta$, i.e., $\dot Y(0)=\Delta$.  Applying this
definition to $F(Y)$ and Eq.~(\ref{eq:sgcomp}) yields the formula
\begin{eqnarray} \Hess F(\Delta_1,\Delta_2) &=& F_{YY}(\Delta_1,\Delta_2) 
+\half\tr\bigl((F_Y\T\Delta_1Y^T+Y\T\Delta_1F_Y^T)\Delta_2\bigr)
\label{eq:Vhess}\\
&&\quad-\half\tr\bigl((Y^TF_Y+F_Y^TY)\Delta_1^T\Pi\Delta_2\bigr),
\nonumber\end{eqnarray} where $\Pi=I-YY^T$, $F_Y$ is defined in
Eq.~(\ref{eq:FYdef}), and the notation $F_{YY}(\Delta_1,\Delta_2)$
denotes the scalar $\sum_{ij,kl}
(F_{YY})_{ij,kl}(\Delta_1)_{ij}(\Delta_2)_{kl}$, where \begin{equation}
(F_{YY})_{ij,kl}={\partial^2 F\over \partial Y_{ij}\partial Y_{kl}}.
\label{eq:FYYdef}\end{equation} This formula may also readily be
obtained by using Eq.~(\ref{polo}) and the formula
\begin{equation} \Hess F(\Delta_1,\Delta_2)=F_{YY}(\Delta_1,\Delta_2)-
\tr F_Y^T  \Gamma_c(\Delta_1,\Delta_2). \label{eq:hess2dcov} \end{equation}

For Newton's method, we must determine the tangent vector $\Delta$
such that \begin{equation}\Hess F(\Delta,X)=\langle -G,X\rangle 
\quad\hbox{for all
tangent vectors $X$,} \label{eq:Hess-1G}\end{equation} where $G=\nabla F$.
Recall that $\langle\,{,}\,\rangle\equiv g_c(\,{,}\,)$ in this
context.  We shall express the solution to this linear equation as
$\Delta=-\Hess^{-1}G$, which may be expressed as the solution to the
linear problem \begin{equation} F_{YY}(\Delta) - Y \skew(F_Y \T \Delta)
-\skew(\Delta F_Y^T ) Y - \frac{1}{2} \Pi \Delta Y^T F_Y=-G,
\label{eq:VHess-1G}\end{equation} $Y\T\Delta={}$skew-symmetric, where
$\skew(X)=(X-X^T)/2$ and the notation $F_{YY}(\Delta)$ means the
unique tangent vector satisfying the equation \begin{equation}
  F_{YY} (\Delta,X) =\langle F_{YY}(\Delta),X\rangle
  \quad\hbox{for all tangent vectors
    $X$.}\label{eq:FYYD}\end{equation}

Example problems are considered in \S{}\ref{sec:geomalg}.

\subsection{The Grassmann Manifold with its Canonical Metric}
\label{sec:grassgeom}

A quotient space representation of the Grassmann manifold was
given in \S{}\ref{sec:qggrass}; however, for computations we
prefer to work with $n\by p$ orthonormal matrices $Y$.  When
performing computations on the Grassmann manifold, we will use
the $n\by p$ matrix $Y$ to represent an entire equivalence class
\begin{equation} [Y]=\{\,YQ_p:Q_p\in O_p\,\},
  \label{eq:eqcgrass}\end{equation} i.e., the subspace spanned by
the columns of~$Y$.  Any representative of the equivalence class
will do.

We remark that an alternative strategy is to represent points on the
Grassmann manifold with projection matrices $YY^T$.  There is one
unique such matrix corresponding to each point on the Grassmann
manifold.  On first thought it may seem foolish to use $n^2$
parameters to represent a point on the Grassmann manifold (which has
dimension $p(n-p)$), but in certain {\it ab initio\/} physics computations
\cite{goedecker95}, the projection matrices $YY^T$ that arise in
practice tend to require only $O(n)$ parameters for their
representation.

Returning to the $n\by p$ representation of points on the Grassmann
manifold, the tangent space is easily computed by viewing the
Grassmann manifold as the quotient space $G_{n,p}=V_{n,p}/O_p$.  At a
point $Y$ on the Stiefel manifold then, as seen in Eq.~(\ref{eq:TV}),
tangent vectors take the form $\Delta =YA+\Yperp B$, where $A$ is
$p\by p$ skew-symmetric, $B$ is $(n-p)\by p$, and $\Yperp$ is any
$n\by(n-p)$ matrix such that $(Y\matsep\Yperp)$ is orthogonal.  From
Eq.~(\ref{eq:eqcgrass}) it is clear that the vertical space at~$Y$
is the set of vectors of the form \begin{equation}\Phi=YA;
\end{equation} therefore, the
horizontal space at~$Y$ is the set of vectors of the form
\begin{equation}\Delta=\Yperp B.\label{eq:TG1}\end{equation}  
Because the horizontal space is equivalent to the tangent space
of the quotient, the tangent space of the Grassmann manifold
at~$[Y]$ is given by all $n\by p$ matrices $\Delta$ of the form
in Eq.~(\ref{eq:TG1}) or, equivalently, all $n\by p$ matrices
$\Delta$ such that \begin{equation}
  Y\T\Delta=0.\label{eq:TG2}\end{equation} Physically, this
corresponds to directions free of rotations mixing the basis
given by the columns of~$Y$.

We already saw in \S{}\ref{sec:qggrass} that the Euclidean metric
is in fact equivalent to the canonical metric for the Grassmann
manifold.  That is, for $n\by p$ matrices $\Delta_1$ and $\Delta_2$
such that $Y\T\Delta_i=0$ ($i=1$, $2$),
\begin{eqnarray*} g_c(\Delta_1,\Delta_2) &=&\tr\Delta_1^T(I-\half
YY^T)\Delta_2,\\ &=&\tr\Delta_1\T\Delta_2,\\
&=&g_e(\Delta_1,\Delta_2).\end{eqnarray*}

\subsubsection{Geodesics (Grassmann)}

A formula for geodesics on the Grassmann manifold was\break 
given 
via
Eq.~(\ref{eq:grassgeod}); the following theorem provides a useful
method for computing this formula using $n\by p$ matrices.

\begin{theorem} 
If $Y(t)=Qe^{t\left({0\atop B}\;{-\smash{B^T}\atop0}\right)}
  I_{n,p}$, with $Y(0)=Y$ and $\Yd(0)=H,$ then \begin{equation}
\label{eq:csgeos}
Y(t) =\pmatrix{YV&U\cr} \pmatrix{\cos\Sigma t\cr\sin\Sigma
  t\cr}V^T, \end{equation} where $U\Sigma V^T$ is the compact
singular value decomposition of $H$.
\end{theorem}

{\em Proof 1}.
It is easy to check that either formulation for the geodesic satisfies
the geodesic equation $\ddot{Y}+Y(\dot Y^T\dot Y)=0$, with the same
initial conditions.
\qquad\endproof

{\em Proof 2}. Let $B =(U_1\matsep
U_2){\Sigma\atopwithdelims()0}V^T$ be the singular value
decomposition of~$B$ ($U_1$ $n\by p$, $U_2$ $p\by(n-p)$, $\Sigma$ and
$V$ $p\by p$).  A straightforward computation involving the
partitioned matrix \begin{equation}\pmatrix{0&-B^T\cr B&0\cr}
=\pmatrix{V&0&0\cr0&U_1&U_2\cr}
\pmatrix{0&-\Sigma&0\cr\Sigma&0&0\cr0&0&0\cr}
\pmatrix{V^T&0\cr0&U_1^T\cr0&U_2^T\cr} \label{eq:TGdecomp}\end{equation}
verifies the theorem.
\qquad\endproof

A subtle point in Eq.~(\ref{eq:csgeos}) is that if the rightmost $V^T$
is omitted, then we still have a representative of the same
equivalence class as~$Y(t)$; however, due to consistency conditions
along the equivalent class $[Y(t)]$, the tangent (horizontal) vectors
that we use for computations must be altered in the same way.  This
amounts to post-multiplying everything by~$V$, or for that matter, any
$p\by p$ orthogonal matrix.

The path length between $Y_0$ and $Y(t)$ (distance between subspaces)
is given by \cite{wong67} \begin{equation} d\bigl(Y(t),Y_0\bigr) =t\|H\|_F
=t\biggl(\sum_{i=1}^p\sigma_i^2\biggr)^{1/2},\label{eq:arclengthd}
\end{equation}
where $\sigma_i$ are the diagonal elements of $\Sigma$. (Actually,
this is only true for $t$ small enough to avoid the issue of conjugate
points, e.g., long great circle routes on the sphere.)  An
interpretation of this formula in terms of the CS decomposition and
principal angles between subspaces is given in
\S{}\ref{sec:csdecomp}.


\subsubsection{Parallel Translation (Grassmann)}

A formula for parallel translation along geodesics of complexity
$O(np^2)$ can also be derived:

\begin{theorem} Let $H$ and $\Delta$ be tangent vectors to the
Grassmann manifold at~$Y$.  Then the parallel translation
of~$\Delta$ along the geodesic in the direction $\Yd(0)=H$
[Eq.~(\ref{eq:csgeos})] is  \begin{equation}
\tau\Delta(t) =\left(\pmatrix{YV&U\cr} \pmatrix{-\sin\Sigma t\cr
\cos\Sigma t}U^T +(I-UU^T)\right)\Delta. \label{eq:ptgrass} \end{equation}
\end{theorem}

{\em Proof 1}.
A simple  computation verifies that Eqs.\ (\ref{eq:ptgrass}) and
(\ref{eq:csgeos}) satisfy \linebreak Eq.~(\ref{eq:ptstiefeuc}).
\qquad\endproof

{\em Proof 2}.
Parallel translation of $\Delta$ is given by the expression
$$\tau\Delta(t) =Q\exp t\pmatrix{0&-A^T\cr A&0\cr}\pmatrix{0\cr
B\cr}$$ (which follows from Eq.~(\ref{eq:ptgrassq})), where
$Q=(Y\matsep\Yperp)$, $H=\Yperp A$, and $\Delta=\Yperp B$.  Decomposing
$\left({0\atop A}\;{-\smash{A^T}\atop0}\right)$ as in Eq.~(\ref{eq:TGdecomp})
(note well that $A$ has replaced $B$), a straightforward computation
verifies the theorem.
\qquad\endproof


\subsubsection{The Gradient of a Function (Grassmann)}
\label{sec:Ggrad}

We must compute the gradient of a function $F(Y)$ defined on the
Grassmann manifold.  Similarly to \S{}\ref{sec:Vgrad}, the
gradient of~$F$ at~$[Y]$ is defined to be the tangent vector
$\nabla F$ such that \begin{equation}\tr F_Y\T\Delta =g_c(\nabla
  F,\Delta) \equiv\tr(\nabla F)^T\Delta
  \label{eq:grassgraddef}\end{equation} for~all tangent vectors
$\Delta$ at~$Y$, where $F_Y$ is defined by Eq.~(\ref{eq:FYdef}).
Solving Eq.~(\ref{eq:grassgraddef}) for $\nabla F$ such that
$Y^T(\nabla F)=0$ yields \begin{equation} \nabla F = F_Y-YY^TF_Y.
  \label{eq:Gnpgradf}
\end{equation} Eq.~(\ref{eq:Gnpgradf}) may also be derived by differentiating
$F\bigl(Y(t)\bigr)$, where $Y(t)$ is the Grassmann geodesic given by
Eq.~(\ref{eq:csgeos}).

\subsubsection{The Hessian of a Function (Grassmann)}
\label{sec:GHess}

Applying the definition for the Hessian of~$F(Y)$ given by
Eq.~(\ref{eq:Hessgdef}) in the context of the Grassmann manifold yields
the formula \begin{equation} \Hess F(\Delta_1,\Delta_2) =
F_{YY}(\Delta_1,\Delta_2) -\tr(\Delta_1^T\Delta_2 Y\T
F_Y),\end{equation} where $F_Y$ and $F_{YY}$ are defined in
\S{}\ref{sec:VHess}.  For Newton's method, we must determine
$\Delta=-\Hess^{-1}G$ satisfying Eq.~(\ref{eq:Hess-1G}), which for the
Grassmann manifold is expressed as the linear problem \begin{equation}
F_{YY}(\Delta) -\Delta(Y^TF_Y)=-G, \label{eq:GHess-1G}\end{equation}
$Y\T\Delta=0$, where $F_{YY}(\Delta)$ denotes the unique tangent
vector satisfying Eq.~(\ref{eq:FYYD}) for the Grassmann manifold's
canonical metric.

Example problems are considered in \S{}\ref{sec:geomalg}.

\subsection{Conjugate Gradient on Riemannian Manifolds}
\label{sec:cgnmrm}

As demonstrated by Smith \cite{smith93a,smith94a}, the benefits of
using the conjugate gradient algorithm for unconstrained minimization
can be carried over to minimization problems constrained to Riemannian
manifolds by a covariant translation of the familiar operations of
computing gradients, performing line searches, the computation of
Hessians, and carrying vector information from step to step in the
minimization process.  In this section we will review the ideas in
\cite{smith93a,smith94a} and then in the next section we formulate
concrete algorithms for conjugate gradient on the Stiefel and
Grassmann manifolds.  Here one can see how the geometry provides
insight into the true difference among the various formulas that are
used in linear and nonlinear conjugate gradient algorithms.

\begin{figure}[pht]
\centerline{\epsfbox{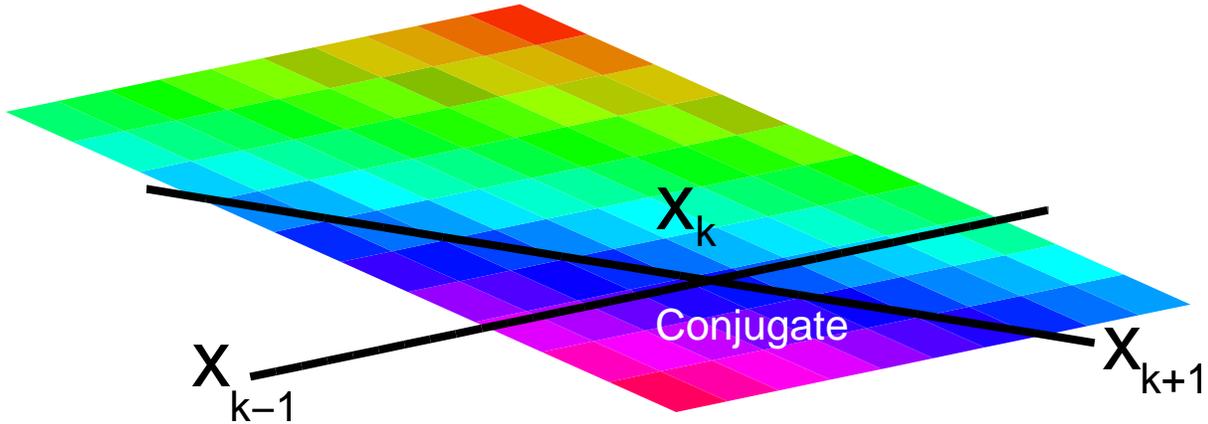}}
\caption{Conjugate gradient in flat space.}
\label{fig:flat}
\end{figure}

\begin{figure}[pht]
\centerline{\epsfbox{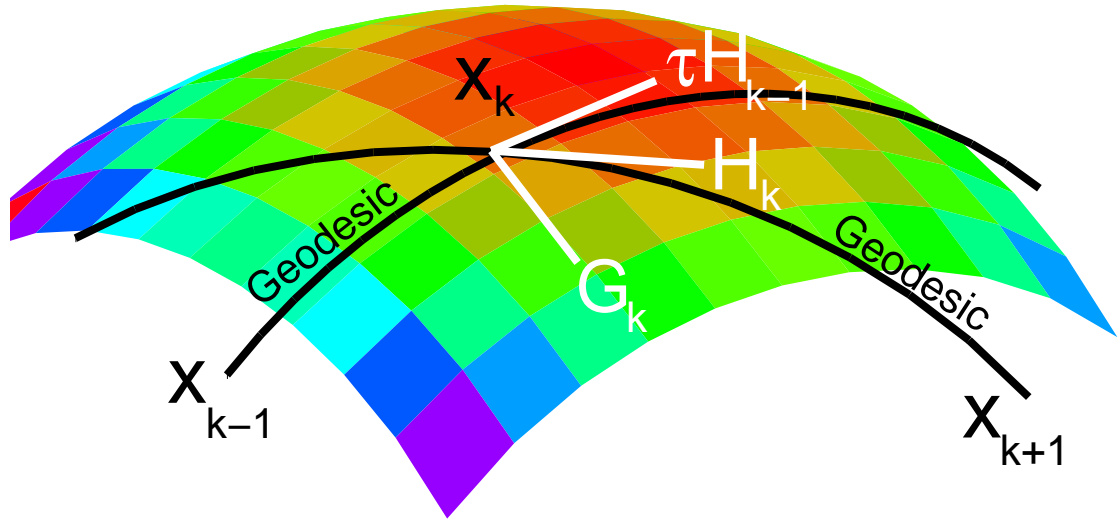}}
\caption{Conjugate gradient in curved space.}
\label{fig:curved}
\end{figure}

Figure \ref{fig:flat} sketches the conjugate gradient algorithm in
flat space and Figure \ref{fig:curved} illustrates the algorithm on a
curved space.  An outline for the iterative part of the algorithm (in
either flat or curved space) goes as follows: at the $(k-1)$st iterate
$x_{k-1}$, step to $x_k$, the minimum of~$f$  along the
geodesic in the direction $H_{k-1}$, compute the gradient
$G_k=\nabla\!f(x_k)$ at this point, choose the new search direction to
be a combination of the old search direction and the new gradient:
\begin{equation} 
H_k=G_k+\gamma_{k}\tau H_{k-1}, \label{eq:cgsd}
\end{equation} 
and iterate until convergence.  Note that $\tau H_{k-1}$ in
Eq.~(\ref{eq:cgsd}) is the parallel translation of the vector
$H_{k-1}$ defined in \S{}\ref{sec:xport}, which in this case is
simply the direction of the geodesic (line) at the point~$x_k$ (see
Figure~\ref{fig:curved}).  Also note the important condition that
$x_k$ is a minimum point along the geodesic:
\begin{equation} \langle G_k, \tau H_{k-1} \rangle = 0 . 
\label{mincond} \end{equation}
\goodbreak

Let us begin our examination of the choice of $\gamma_k$ in flat space
before proceeding to arbitrary manifolds. Here parallel transport is
trivial so that $$H_k = G_k + \gamma_{k} H_{k-1}.$$ In both linear and
an idealized version of nonlinear conjugate gradient, $\gamma_{k}$ may
be determined by the exact conjugacy condition for the new search
direction: $$f_{xx}(H_{k},H_{k-1})=0,$$ i.e., the old and new search
direction must be conjugate with respect to the Hessian of $f$.  (With
$f_{xx}=A$, the common notation \cite[page 523]{golub89a} for the
conjugacy condition is $p_{k-1}^T \! A p_k=0$.)  The formula for
$\gamma_k$ is then \begin{eqnarray} \label{exc}
\hbox to4pc{\hss Exact Conjugacy:\quad}\gamma_k &=& -
f_{xx}(G_k, H_{k-1})/f_{xx}(H_{k-1}, H_{k-1})  .
\end{eqnarray}

The standard trick to improve the computational efficiency of linear
conjugate gradient is to use a formula relating a finite difference of
gradients to the Hessian times the direction ($r_k-r_{k-1}=-\alpha_k A
p_k$ as in \cite{golub89a}).  In our notation, \begin{equation}
\langle G_k-G_{k-1},\cdot\rangle \approx \alpha f_{xx}(\cdot, H_{k-1}),
\label{hessc} \end{equation} where $\alpha=\|x_k-x_{k-1}\|/\|H_{k-1}\|$.

The formula is exact for linear conjugate gradient on flat space,
otherwise it has the usual error in finite difference approximations.
By applying the finite difference formula Eq.~(\ref{hessc}) in both
the numerator and denominator of Eq.~(\ref{exc}), and also applying
Eq.~(\ref{mincond}) twice (once with $k$ and once with $k-1$), one
obtains the formula \begin{eqnarray} \hbox to4pc{\hss
Polak-Ribi\`ere:\quad}\gamma_k &=&
\langle G_{k}- G_{k-1},G_k\rangle/\langle
G_{k-1},G_{k-1}\rangle. \end{eqnarray} Therefore the Polak-Ribi\'ere
formula is the exact formula for conjugacy through the Hessian, where
one uses a difference of gradients as a finite difference
approximation to the second derivative.  If $f(x)$ is well
approximated by a quadratic function, then $\langle G_{k-1}, G_k
\rangle \approx 0$, and we obtain \begin{eqnarray}
\hbox to4pc{\hss Fletcher-Reeves:\quad}\gamma_k &=& \langle
G_{k},G_k\rangle/\langle G_{k-1},G_{k-1}\rangle. \end{eqnarray}

For arbitrary manifolds, the Hessian is the second derivative along
geodesics.  In differential geometry it is the second covariant
differential of $f$.  Here are the formulas:
\begin{eqnarray}
\hbox to10em{\hss Exact Conjugacy:\quad}\gamma_k &=& -
      \Hess\!f(G_k,\tau H_{k-1})/\Hess\!f(\tau H_{k-1},\tau H_{k-1}) \\
\hbox to10em{\hss Polak-Ribi\`ere:\quad}\gamma_k &=& \langle G_{k}-\tau
      G_{k-1},G_k\rangle/\langle G_{k-1},G_{k-1}\rangle\\ 
\hbox to10em{\hss Fletcher-Reeves:\quad}\gamma_k &=& \langle G_{k},G_k\rangle/\langle
      G_{k-1},G_{k-1}\rangle
\end{eqnarray} 
which may be derived from the
finite difference approximation to the Hessian,
$$
\langle G_k -\tau G_{k-1},\cdot\rangle \approx \alpha
\Hess\!f(\cdot,\tau H_{k-1}),\quad
\alpha=d(x_k,x_{k-1})/\|H_{k-1}\|.$$

Asymptotic analyses appear in \S{}\ref{sec:approx}.

\section{Geometric Optimization Algorithms}
\label{sec:geomalg}

The algorithms presented here are our answer to the question: ``What
does it mean to perform the Newton and conjugate gradient methods on
the Stiefel and Grassmann manifolds?''  Though these algorithms are
idealized, they are of identical complexity up to small constant
factors with the best known algorithms.  In particular, no
differential equation routines are used. It is our hope that in the
geometrical algorithms presented here, the reader will recognize
elements of any algorithm that accounts for orthogonality constraints.
These algorithms are special cases of the Newton and conjugate
gradient methods on general Riemannian manifolds.  If the objective
function is nondegenerate, then the algorithms are guaranteed to
converge quadratically \cite{smith93a,smith94a}.

\clearpage
\subsection{Newton's Method on the Grassmann Manifold}
\label{sec:nmgrass}

In flat space, Newton's method simply updates a vector by subtracting
the gradient vector pre-multi\-plied by the inverse of the Hessian.  The
same is true on the Grassmann manifold (or any Riemannian manifold for
that matter) of $p$-planes in $n$-dimensions, with interesting
modifications.  Subtraction is replaced by following a geodesic path.
The gradient is the usual one (which must be tangent to the constraint
surface), and the Hessian is obtained by twice differentiating the
function along a geodesic.  We show in \S{}\ref{sec:lagrange} that
this Hessian is related to the Hessian of the Lagrangian; the two
Hessians arise from the difference between the intrinsic and extrinsic
viewpoints.  It may be suspected that following geodesics may not be
computationally feasible, but because we exploit the structure of the
constraint surface, this operation costs $O(np^2)$, which is required
even for traditional algorithms for the eigenvalue problem---our
simplest example.

Let $F(Y)$ be a smooth function on the Grassmann manifold, i.e.,
$F(Y)=F(YQ)$ for any $p\by p$ orthogonal matrix $Q$, where $Y$ is an
$n\by p$ matrix such that $Y^TY=I_p$.  We compute formulas for $F_Y$ and
$F_{YY}(\Delta)$ using the definitions given in
\S{}\ref{sec:GHess}.  Newton's method for minimizing $F(Y)$ on the
Grassmann manifold is:

\bigskip
\masbox{ {\sc Newton's Method for Minimizing
$F(Y)$ on the Grassmann Manifold}\\
\stepitem Given $Y$ such that $Y^TY=I_p$,
\foritem Compute $G=F_Y-YY^TF_Y$.
\foritem Compute $\Delta=-\Hess^{-1}G$ such that
      $Y\T\Delta=0$ and $$F_{YY}(\Delta)-\Delta(Y^TF_Y)=-G.$$
\stepitem Move from $Y$ in direction $\Delta$ to~$Y(1)$ using the geodesic
      formula $$Y(t) =YV\cos(\Sigma t)V^T + U\sin(\Sigma t)V^T$$ where 
      $U\Sigma V^T$ is the compact singular
      value decomposition of $\Delta$ (meaning $U$ is $n \by p$ and both
      $\Sigma$ and $V$ are $p \by p$).
\stepitem Repeat.
}
\bigskip

The special case of minimizing $F(Y)=\frac{1}{2}\tr Y\T AY$ ($A$ $n\by
n$ symmetric) gives the geometrically correct Newton method for the
symmetric eigenvalue problem.  In this case $F_Y=AY$ and
$F_{YY}(\Delta)=(I-YY^T)A\Delta$.  The resulting algorithm requires the
solution of a Sylvester equation. It is the idealized algorithm whose
approximations include various forms of Rayleigh quotient iteration,
inverse iteration, a number of Newton style methods for invariant
subspace computation, and the many variations of Davidson's eigenvalue
method.  These ideas are discussed in \S{}\ref{sec:rqi} and
\ref{sec:nmisc}. 

\clearpage
\subsection{Newton's Method on the Stiefel Manifold}
\label{sec:nmstief}

Newton's method on the Stiefel manifold is conceptually equivalent to
the Grassmann manifold case.  Let $Y$ be an $n\by p$ matrix such that
$Y^TY=I_p$, and let $F(Y)$ be a smooth function of~$Y$, without the
homogeneity condition imposed for the Grassmann manifold case.
Compute formulas for $F_Y$ and $F_{YY}(\Delta)$ using the definitions
given in \S{}\ref{sec:VHess}.  Newton's method for minimizing
$F(Y)$ on the Stiefel manifold is:

\bigskip
\masbox{
{\sc Newton's Method for Minimizing $F(Y)$ on the Stiefel Manifold} \\
\stepitem Given $Y$ such that $Y^TY=I_p$,
\foritem Compute $G=F_Y-YF_Y^TY$.
\foritem Compute $\Delta=-\Hess^{-1}G$ such that
      $Y\T\Delta={}$skew-symmetric and $$F_{YY}(\Delta)
      -Y \skew(F_Y \T \Delta) -\skew(\Delta F_Y^T)Y-\half \Pi\Delta Y^T
      F_Y=-G,$$ where $\skew(X)=(X-X^T)/2$ and $\Pi=I-YY^T$.
\stepitem Move from $Y$ in direction $\Delta$ to~$Y(1)$ using the geodesic
      formula $$Y(t)=YM(t)+QN(t)$$ where $QR$ is the compact QR
      decomposition of $(I-YY^T)\Delta$ (meaning $Q$ is $n \by p$ and
      $R$ is $p\by p$), $A=Y\T\Delta$, and $M(t)$ and $N(t)$ are $p\by
      p$ matrices given by the $2p\by 2p$ matrix exponential
      $$\pmatrix{M(t)\cr N(t)\cr}=\exp t\pmatrix{A&-R^T\cr R&0\cr}
      \pmatrix{I_p\cr 0\cr}.$$
\stepitem Repeat.
}
\bigskip

For the special case of minimizing $F(Y)=\frac{1}{2}\tr Y\T AYN$ ($A$
$n\by n$ symmetric, $N$ $p\by p$ symmetric) \cite{smith93a}, $F_Y=AYN$
and $F_{YY}(\Delta)= A\Delta N-YN\Delta\T A Y$.  Note that if $N$ is
not a multiple of the identity, then $F(Y)$ does not have the
homogeneity condition required for a problem on the Grassmann
manifold.  If $N=\diag(p,p-1,\ldots,1)$, then the optimum solution to
maximizing $F$ over the Stiefel manifold yields the eigenvectors
corresponding to the $p$ largest eigenvalues.

For the orthogonal Procrustes problem \cite{elden77},
$F(Y)={1\over2}\|AY-B\|_F^2$ ($A$ $m\by n$, $B$ $m\by p$, both
arbitrary), $F_Y=A\T AY-A^TB$ and $F_{YY}(\Delta)= A\T
A\Delta-Y\Delta\T A\T AY$.  Note that $Y^TF_{YY}(\Delta)
={}$skew-symmetric.

\clearpage
\subsection{Conjugate Gradient Method on the Grassmann Manifold}
\label{sec:cggrass}

Conjugate gradient techniques are considered because they are easy to
implement, have low storage requirements, and provide superlinear
convergence in the limit.  The Newton equations may be solved with
finitely many steps of linear conjugate gradient; each nonlinear
conjugate gradient step, then, approximates a Newton step.  In flat
space, the nonlinear conjugate gradient method performs a line search
by following a direction determined by conjugacy with respect to the
Hessian.  On Riemannian manifolds, conjugate gradient performs
minimization along geodesics with search directions defined using the
Hessian described above \cite{smith93a,smith94a}.  Both algorithms
approximate Hessian conjugacy with a subtle formula involving only the
gradient directions, resulting in an algorithm that captures second
derivative information by computing only first derivatives.
To ``communicate'' information from one iteration to the next, tangent
vectors must parallel transport along geodesics.  Conceptually, this
is necessary because, unlike flat space, the definition of tangent
vectors changes from point to point.  

\samepage{Using these ideas and formulas developed in \S{}\ref{sec:nmgrass},
the conjugate gradient method on the Grassmann manifold is:
}

\bigskip
\masbox{
{\sc Conjugate Gradient for Minimizing $F(Y)$ on the Grassmann Manifold}
\stepitem Given $Y_0$ such~that $Y_0^TY_0=I$, compute
      $G_0=F_{Y_0}-Y_0Y_0^TF_{Y_0}$ and set $H_0=-G_0$.
\stepitem For $k=0$, $1$, \dots,
\foritem Minimize $F\bigl(Y_k(t)\bigr)$ over~$t$ where $$Y(t)
      =YV\cos(\Sigma t)V^T + U\sin(\Sigma t)V^T$$ and $U\Sigma V^T$ is
      the compact singular value decomposition of $H_k$.
\foritem Set $t_k=t_{\rm min}$ and $Y_{k+1}=Y_k(t_k)$.
\foritem Compute $G_{k+1}=F_{Y_{k+1}}-Y_{k+1}Y_{k+1}^TF_{Y_{k+1}}$.
\foritem Parallel transport tangent vectors $H_k$ and $G_k$
      to the point $Y_{k+1}$: \begin{eqnarray} \tau H_k
      &=&(-Y_kV\sin\Sigma t_k +U\cos\Sigma t_k)\Sigma
      V^T,\label{eq:GnptH}\\ \tau G_k &=&G_k-\bigl(Y_kV\sin\Sigma t_k
      +U(I-\cos\Sigma t_k)\bigr)U\T G_k. \label{eq:GnptG}\end{eqnarray}
\foritem Compute the new search direction $$H_{k+1}
      =-G_{k+1} +\gamma_k\tau H_k\qquad\hbox{where}\qquad \gamma_k = {\langle
      G_{k+1} -\tau G_k,G_{k+1}\rangle\over \langle G_k,G_k\rangle}$$
      and $\langle \Delta_1,\Delta_2\rangle=\tr \Delta_1\T\Delta_2$.
\foritem Reset $H_{k+1}=-G_{k+1}$ if $k+1\equiv0\bmod{p(n-p)}$.
}
\bigskip

\subsection{Conjugate Gradient Method on the Stiefel Manifold}
\label{sec:cgstief}

As with Newton's method, conjugate gradient on the two manifolds is
very similar.  One need only replace the definitions of tangent
vectors, inner products, geodesics, gradients, and parallel
translation.  Geodesics, gradients, and inner products on the Stiefel
manifold are given in \S{}\ref{sec:stiefgeom}.  For parallel
translation along geodesics on the Stiefel manifold, we have no
simple, general formula comparable to Eq.~(\ref{eq:GnptG}).
Fortunately, a geodesic's tangent direction is parallel, so a simple
formula for $\tau H_k$ comparable to Eq.~(\ref{eq:GnptH}) is
available, but a formula for $\tau G_k$ is not.  In practice, we
recommend setting $\tau G_k:=G_k$ and ignoring the fact that $\tau
G_k$ will not be tangent at the point $Y_{k+1}$.  Alternatively,
setting $\tau G_k:=0$ (also not parallel) results in a Fletcher-Reeves
conjugate gradient formulation.  As discussed in the next section,
either approximation does not affect the superlinear convergence
property of the conjugate gradient method.

The conjugate gradient method on the Stiefel manifold is:

\bigskip
\masbox{
{\sc Conjugate Gradient for Minimizing $F(Y)$ on the Stiefel Manifold}
\stepitem Given $Y_0$ such~that $Y_0^TY_0=I$, compute $G_0
      =F_{Y_0}-Y_0F_{Y_0}^TY_0$ and set $H_0=-G_0$.
\stepitem For $k=0$, $1$, \dots,
\foritem Minimize $F\bigl(Y_k(t)\bigr)$ over~$t$ where
      $$Y_k(t)=Y_kM(t)+QN(t),$$ $QR$ is the compact QR decomposition
      of $(I-Y_kY_k^T)H_k$, $A=Y_k^TH_k$, and $M(t)$ and $N(t)$ are
      $p\by p$ matrices given by the $2p\by 2p$ matrix exponential
      appearing in Newton's method on the Stiefel manifold in
      \S{}\ref{sec:nmstief}.
\foritem Set $t_k=t_{\rm min}$ and $Y_{k+1}=Y_k(t_k)$.
\foritem Compute $G_{k+1} =F_{Y_{k+1}}-Y_{k+1}F_{Y_{k+1}}^TY_{k+1}$.
\foritem Parallel transport tangent vector $H_k$ to the point
      $Y_{k+1}$: \begin{equation}\tau H_k
        =H_kM(t_k)-Y_kR^TN(t_k).\label{eq:sptcomp}
\end{equation} As discussed
      above, set $\tau G_k:=G_k$ or~$0$, which is {\it not\/} parallel.
\foritem Compute the new search direction $$H_{k+1} =-G_{k+1}
      +\gamma_k\tau H_k\qquad\hbox{where}\qquad \gamma_k = {\langle
      G_{k+1} -\tau G_k,G_{k+1}\rangle\over \langle G_k,G_k\rangle}$$
      and $\langle \Delta_1,\Delta_2\rangle=\tr
      \Delta_1^T(I-{1\over2}YY^T)\Delta_2$.
\foritem Reset $H_{k+1}=-G_{k+1}$ if $k+1\equiv0\bmod{p(n-p)
      +p(p-1)/2}$.
}
\bigskip

\clearpage
\subsection{Numerical Results and Asymptotic Behavior}
\label{sec:numres}

\subsubsection{Trace Maximization on the Grassmann Manifold}
\label{sec:trmin}

\begin{figure}[hbt]
\centerline{\epsfysize=3in\epsfbox[148 261 484 522]{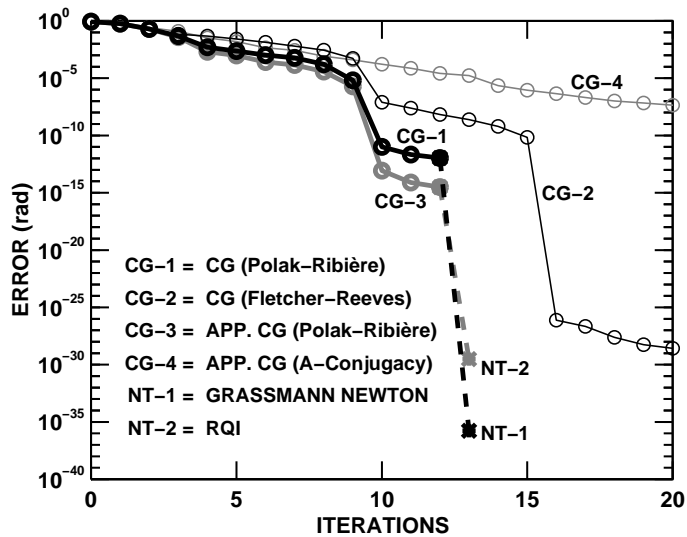}}
\caption[Numerical Results]{Convergence of the conjugate gradient and
Newton's method for trace maximization on the Grassmann manifold
$G_{5,3}$.  The error (in radians) is the arc~length distance between
the solution and the subspace at the $i$th iterate
[Eq.~(\ref{eq:arclengthd}) and \S{}\ref{sec:csdecomp}].  Quadratic
convergence of CG is evident, as is cubic convergence of Newton's
method, which is a special property of this example.}
\label{fig:numres}
\end{figure}

The convergence properties of the conjugate gradient and Newton's
methods applied to the trace maximization problem $F(Y)=\tr Y\T AY$
are shown in Figure~\ref{fig:numres}, as well as the convergence of an
approximate conjugate gradient method and the Rayleigh quotient
iteration for comparison.  This example shows trace maximization on
$G_{5,3}$, i.e., $3$-dimensional subspaces in $5$ dimensions.  The
distance between the subspace and the known optimum subspace is
plotted versus the iteration number, where the distance in radians is
simply the square root of the sum of squares of the principal angles
between the subspaces.  The dimension of this space equals $3(5-3)=6$;
therefore, a conjugate gradient algorithm with resets should at least
double in accuracy every six iterations.  Newton's method, which is
cubically convergent for this example (this point is discussed in
\S{}\ref{sec:rqi}), should triple in accuracy every iteration.
Variable precision numerical software is used to demonstrate the
asymptotic convergence properties of these algorithms.

The thick black curve (CG-1) shows the convergence of the conjugate
gradient (CG) algorithm using the Polak-Ribi\`ere formula.  The
accuracy of this algorithm is at least doubled between the first and
sixth and the seventh and twelfth iterations, demonstrating this
method's superlinear convergence.  Newton's method is applied to the
twelfth CG iterate, which results in a tripling of the accuracy and
demonstrates cubic convergence of Newton's method, shown by the dashed
thick black curve (NT-1).

The thin black curve (CG-2) shows CG convergence using the
Fletcher-Reeves formula 
\begin{equation}\gamma_k=\langle
G_{k+1},G_{k+1}\rangle/\langle G_k,G_k\rangle.
\end{equation} 
\goodbreak
As discussed below,
this formula differs from the Polak-Ribi\`ere formula by second order
and higher terms, so it must also have superlinear convergence.  The
accuracy of this algorithm more than doubles between the first and
sixth, seventh and twelfth, and thirteenth and eighteenth iterations,
demonstrating this fact.

\goodbreak
The algorithms discussed above are actually performed on the
constraint surface, but extrinsic approximations to these algorithms
are certainly possible.  By perturbation analysis of the metric given
below, it can be shown that the CG method differs from its flat space
counterpart only by cubic and higher terms close to the solution;
therefore, a flat space CG method modified by projecting search
directions to the constraint's tangent space will converge
superlinearly.  This is basically the method proposed by Bradbury and
Fletcher \cite{bradbury66a} and others for the single eigenvector case.  For
the Grassmann (invariant subspace) case, we have performed line
searches of the function $\phi(t)=\tr Q(t)\T AQ(t)$, where
$Q(t)R(t):=Y+t\Delta$ is the compact QR decomposition, and
$Y\T\Delta=0$. The QR decomposition projects the solution back to the
constraint surface at every iteration.  Tangency of the search
direction at the new point is imposed via the projection ${I-YY^T}$.

The thick gray curve (CG-3) illustrates the superlinear convergence of
this method when the Polak-Ribi\`ere formula is used.  The
Fletcher-Reeves formula yields similar results.  In contrast, the thin
gray curve (CG-4) shows convergence when conjugacy through the matrix
$A$ is used, i.e., $\gamma_k=-(H_k\T AG_{k+1})/(H_k\T AH_k)$, which
has been proposed by several authors \cite[Eq.~(5)]{perdon86},
\cite[Eq.~(32)]{chen86}, \cite[Eq.~(20)]{fu95}.  This method cannot be
expected to converge superlinearly because the matrix $A$ is in fact
quite different from the true Hessian on the constraint surface.  This
issue is discussed further in \S{}\ref{sec:cgeig}.


\begin{figure}[bht]
\centerline{\epsfysize=3in\epsfbox[143 268 474 530]{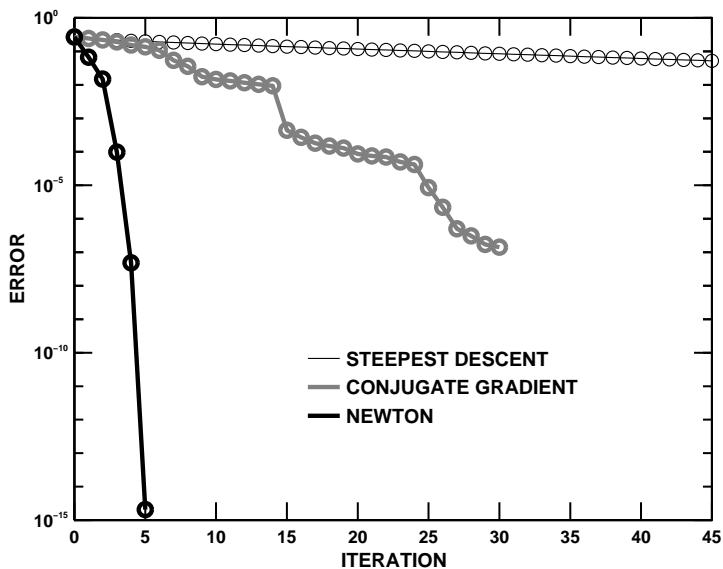}}
\caption[Numerical Results]{Convergence of the conjugate gradient and
Newton's method for the orthogonal Procrustes problem on the Stiefel
manifold $V_{5,3}$.  The error is the Frobenius norm between the $i$th
iterate and the known solution.  Quadratic convergence of the CG and
Newton methods is evident.  The Newton iterates correspond to those of
Table~\protect\ref{tab:procrustes}.}
\label{fig:procrustes}
\end{figure}

\goodbreak
To compare the performance of Newton's method to the Rayleigh quotient
iteration (RQI), which approximates Newton's method to high order (or
vice versa), RQI is applied to the approximate CG method's twelfth
iterate, shown by the dashed thick gray curve (NT-2).

\subsubsection{Orthogonal Procrustes Problem on the Stiefel Manifold}

\begin{table}[t]
\footnotesize
\begin{center}
\vbox{\tabskip=0pt
\halign to\hsize{\hfil$#$\enspace\tabskip=0pt plus1fil
  &\quad$\displaystyle#$&\quad\hfil$\displaystyle#$\tabskip=0pt\cr
\noalign{\hrule\vskip3pt}
\omit\hfil Iterate $i$&\omit\quad\hfil$\|Y_i-\hat Y\|_F$\hfil
  &\omit\quad\hfil$Y_i$\hfil\cr
\noalign{\vskip3pt\hrule\vskip6pt}
0&2.68\times10^{-01}&
\rmatrix{
   0.98341252163956& -0.09749309852408& -0.06630579165572\cr
   0.08482117605077&  0.99248149019173& -0.02619408666845\cr
   0.08655810575052&  0.02896396566088&  0.98816425471159\cr
   0.01388126419090&  0.00902267322408&  0.00728525462855\cr
   0.13423928340551&  0.06749272129685& -0.13563090573981\cr
}\cr \noalign{\vskip6pt}
1&6.71\times10^{-02}&
\rmatrix{
   0.99954707914921&  0.01554828497046&  0.00423211303447\cr
  -0.01656743168179&  0.99905154070826&  0.01216605832969\cr
  -0.00306529752246& -0.01070234416262&  0.99915251911577\cr
  -0.00910501510207& -0.01286811040265&  0.00924631200657\cr
  -0.02321334579158& -0.03706941336228&  0.03798454294671\cr
}\cr \noalign{\vskip6pt}
2&1.49\times10^{-02}&
\rmatrix{
   0.99993878247585&  0.00296823825310&  0.00486487784745\cr
  -0.00301651579786&  0.99998521441661&  0.00192519989544\cr
  -0.00479673956404& -0.00191288709538&  0.99996440819180\cr
  -0.00311307788732& -0.00157358730922&  0.00121316839587\cr
  -0.00897953054292& -0.00382429023234&  0.00650669969719\cr
}\cr \noalign{\vskip6pt}
3&9.77\times10^{-05}&
\rmatrix{
   0.99999999888990&  0.00000730457866& -0.00003211124313\cr
  -0.00000730341460&  0.99999999951242&  0.00000603747062\cr
   0.00003210887572& -0.00000603508216&  0.99999999682824\cr
   0.00000457898008& -0.00001136276061&  0.00002209393458\cr
   0.00003339025497& -0.00002750041840&  0.00006919392999\cr
}\cr \noalign{\vskip6pt}
4&4.81\times10^{-08}&
\rmatrix{
   1.00000000000000&  0.00000000813187&  0.00000001705718\cr
  -0.00000000813187&  1.00000000000000&  0.00000000613007\cr
  -0.00000001705718& -0.00000000613007&  1.00000000000000\cr
  -0.00000001001345& -0.00000000397730&  0.00000000429327\cr
  -0.00000002903373& -0.00000000827864&  0.00000002197399\cr
}\cr \noalign{\vskip6pt}
5&2.07\times10^{-15}&
\rmatrix{
   1.00000000000000&  0.00000000000000&  0.00000000000000\cr
   0.00000000000000&  1.00000000000000&  0.00000000000000\cr
   0.00000000000000&  0.00000000000000&  1.00000000000000\cr
   0.00000000000000&  0.00000000000000&  0.00000000000000\cr
   0.00000000000000&  0.00000000000000&  0.00000000000000\cr
}\cr \noalign{\vskip6pt\hrule\vskip6pt}
\noalign{\centerline{$A=$}\vskip4pt}
\multispan3\hidewidth$\displaystyle\rmatrix{
   0.59792470347241& -1.60148995048070&  1.29611959631725&
   0.00742708895676&    
-0.09653196026400\cr
  -0.34991267564713&  1.03005546700300&  0.38145454055699&
  0.14195063498923&    
-0.16309797180034\cr
   0.16783050038338&  0.51739189509778& -0.42204935150912&
   1.75394028742695&    
-0.63865179066515\cr
   0.24927536521443& -1.34694675520019&  0.92362255783368&
   0.62648865033822&    
-0.31561702752866\cr
  -0.24846337483192& -0.44239067350975& -1.52598136000449&
  0.89515519875598&     
0.87362106204727\cr
}$\cr \noalign{\vskip6pt\hrule}}}
\end{center}
\caption[Newton's method for the Procrustes problem on the Stiefel
manifold.]{Newton's method applied to the orthogonal Procrustes
problem on the Stiefel manifold using the MATLAB code given in this
section.  The matrix $A$ is given below the numerical results, and
$B=AI_{5,3}$.  The quadratic convergence of Newton's method, shown by
the Frobenius norm of the difference between $Y_i$ and $\hat
Y=I_{5,3}$, is evident.  This convergence is illustrated in
Figure~\protect\ref{fig:procrustes}.  It is clear from this example
that the difference $Y_i-\hat Y$ approaches a tangent vector at~$\hat
Y=I_{n,p}$, i.e., $\hat Y^T(Y_i-\hat Y)\to{}$skew-symmetric.}
\label{tab:procrustes}
\end{table}

\goodbreak

The orthogonal Procrustes problem \cite{elden77} 
\begin{equation}\min_{Y\in
V_{n,p}} \|AY-B\|_F\quad\hbox{$A$, $B$ given
matrices,}
\end{equation} 
\goodbreak
is a
minimization problem defined on the Stiefel manifold that has no known
analytical solution for $p$ different from $1$ or $n$.  To ensure that
the objective function is smooth at optimum points, we shall consider
the equivalent problem 
\begin{equation}\min_{Y\in V_{n,p}}
{1\over2}\|AY-B\|_F^2.
\end{equation} 
\goodbreak
Derivatives of this function appear at the end
of \S{}\ref{sec:nmstief}.  MATLAB code for Newton's method applied
to this problem appears below.  Convergence of this algorithm for the
case $V_{5,3}$ and test matrices $A$ and $B$ is illustrated in
Table~\ref{tab:procrustes} and Figure~\ref{fig:procrustes}.  The
quadratic convergence of Newton's method and the conjugate gradient
algorithm is evident.  The dimension of~$V_{5,3}$ equals
$3(3-1)/2+6=9$; therefore, the accuracy of the conjugate gradient
should double every nine iterations, as it is seen to do in
Figure~\ref{fig:procrustes}.  Note that the matrix $B$ is chosen such
that a trivial solution $\hat Y=I_{n,p}$ to this test optimization
problem is known.

\goodbreak

\def\coderule#1{\par\goodbreak\medskip\smallskip
  \hbox to\hsize{\hss\setbox0=\hbox{#1}%
    \dimen0=.618\hsize \advance\dimen0 by-\wd0 \divide\dimen0 by2
    $\vcenter{\hbox{\vrule height.4pt width\dimen0}}$\enspace
    \box0\enspace$\vcenter{\hbox{\vrule height.4pt width\dimen0}}$\hss}
  \penalty10000\smallskip\nointerlineskip}

\vskip12pt
\hrule
\vskip6pt
\centerline{\sc MATLAB Code for Procrustes Problem on the Stiefel Manifold}
\vskip6pt
\hrule
\beginmatlab
n = 5; p = 3;

A = randn(n);
B = A*eye(n,p);
Y0 = eye(n,p);        
H = 0.1*randn(n,p);   H = H - Y0*(H'*Y0);  
Y = stiefgeod(Y0,H);  

d = norm(Y-Y0,'fro')
while d > sqrt(eps)
   Y = stiefgeod(Y,procrnt(Y,A,B));
   d = norm(Y-Y0,'fro')
end
"endmatlab
\coderule{function stiefgeod}
\beginmatlab
function [Yt,Ht] = stiefgeod(Y,H,t)
%

[n,p] = size(Y);

if nargin < 3, t = 1; end
A = Y'*H;  A = (A - A')/2;  
[Q,R] = qr(H - Y*A,0);
MN = expm(t*[A,-R';R,zeros(p)]);  MN = MN(:,1:p);

Yt = Y*MN(1:p,:) + Q*MN(p+1:2*p,:);  
if nargout > 1, Ht = H*MN(1:p,:) - Y*(R'*MN(p+1:2*p,:)); end  
"endmatlab

\coderule{function procrnt}
\beginmatlab
function H = procrnt(Y,A,B)

[n,p] = size(Y);
AA = A'*A;  FY = AA*Y - A'*B;  YFY = Y'*FY;  G = FY - Y*YFY';

dimV = p*(p-1)/2 + p*(n-p);             

H = zeros(size(Y));  R1 = -G;  P = R1;  P0 = zeros(size(Y));
for k=1:dimV
   normR1 = sqrt(stiefip(Y,R1,R1));
   if normR1 < prod(size(Y))*eps, break; end
   if k == 1, beta = 0; else, beta = (normR1/normR0)^2; end
   P0 = P;  P = R1 + beta*P;  FYP = FY'*P;  YP = Y'*P;
   LP = AA*P - Y*(P'*AA*Y) ...    
        - Y*((FYP-FYP')/2) - (P*YFY'-FY*YP')/2 - (P-Y*YP)*(YFY/2);  
   alpha = normR1^2/stiefip(Y,P,LP);  H = H + alpha*P;
   R0 = R1;  normR0 = normR1;  R1 = R1 - alpha*LP;
end
"endmatlab
\coderule{function stief\/ip}
\beginmatlab
function ip = stiefip(Y,A,B)

ip = sum(sum(conj(A).*(B - Y*((Y'*B)/2))));  
"endmatlab
\bigskip
\hrule
\vskip2ex

\subsection{Convergence Rates of Approximate  Methods}
\label{sec:approx}

The algorithms presented in the previous sections are idealized in
that geometrically natural ideas such as geodesics and parallel
translation are used in their definitions.  However, approximations
can yield quadratic rates of convergence.  In the limit, the
Riemannian algorithms approach their Euclidean counterparts in the
tangent plane of the solution point.  A perturbation analysis shows
which terms are necessary and which terms are not necessary to achieve
quadratic convergence.  The following argument holds for any
Riemannian manifold, and therefore applies to either the Grassmann or
Stiefel manifold case.

Consider the CG method applied to a function $F(Y)$ starting at a
point $Y$ within distance $\epsilon$ (small) of the solution $\hat Y$.
For a manifold of dimension~$d$, we must perform a sequence of $d$
steps that take us within distance $O(\epsilon^2)$ of the solution
$\hat Y$.  The Riemannian CG method 
$$\eqalign{H_{\rm new}&=-G_{\rm
new} +\gamma\tau H_{\rm old},\quad \gamma={\langle G_{\rm new}-\tau
G_{\rm old},G_{\rm new}\rangle\over\|G_{\rm old}\|^2};\cr Y_{\rm
new}&=Y(t_{\rm min}),\quad Y(0)=Y_{\rm old},\quad \dot Y(0)=H_{\rm
new};\cr}
$$ 
\goodbreak
\noindent
does this, but we wish to approximate this procedure.
Within a ball of size $O(\epsilon)$ around $\hat Y$, these quantities
have sizes of the following orders:
$$\vbox{\tabskip=.5em\halign
to.4\hsize{$\displaystyle#$\hfil\tabskip=.5em plus1fil
&#\hfil\tabskip=.5em\cr
\noalign{\hrule\vskip3pt}
\omit Order\hfil&\omit Quantity\hfil\cr
\noalign{\vskip3pt\hrule\vskip3pt}
O(1)&$t_{\rm min}$, $\gamma$\cr
O(\epsilon)&$G$, $H$ (new or old)\cr
O(\epsilon^2)&$\|G\|^2$, $\|H\|^2$ (new or old)\cr
O(\epsilon^3)&$\langle \tau G_{\rm old},G_{\rm new}\rangle$\cr
\noalign{\vskip3pt\hrule}}}$$
Also, by perturbation analysis of the Riemannian metric
\cite[Vol.~2, Chap.~4, Props.\ 1 and 6]{chavel93,spivak79}, we have
$$\eqalign{Y(\epsilon)&=Y(0)+\epsilon\Delta+O(\epsilon^3)\cr
\tau G(\epsilon) &=G +O(\epsilon^2)\cr
\langle\,{,}\,\rangle&=I
+O(\epsilon^2)\cr}$$ where $Y(\epsilon)$ is a geodesic in
direction~$\Delta$, $\tau G(\epsilon)$ is parallel translation of~$G$
along~$Y(\epsilon)$, and the last approximation is valid for an
orthonormal basis of the tangent plane at~$Y(\epsilon\Delta)$ and $I$
is the identity.

Inserting these asymptotics into the formulas for the CG method show
that near the solution, eliminating the Riemannian terms gives
$O(\epsilon^3)$ perturbations of the CG sequence, and therefore does
not affect the quadratic rate of convergence.  \hbox{Furthermore,} it can
also be seen that eliminating the Polak-Ribi\`ere term\break
\hbox{$-\langle\tau
G_{\rm old},G_{\rm new}\rangle\big/\|G_{\rm old}\|^2$,} yielding the
Fletcher-Reeves algorithm, perturbs the CG sequence by $O(\epsilon^2)$
terms, which does not affect the quadratic rate of convergence.  Thus
the approximate CG methods discussed in \ref{sec:trmin} converge
quadratically.

\section{Examples: Insights and Applications}
\label{sec:unified}

In this section, we consider ideas from the literature as applications
of the framework and methodology developed in this paper.  It is our
hope that some readers who may be familiar with the algorithms
presented here, will feel that they now really see them with a new
deeper, but ultimately clearer understanding.  It is our further hope
that developers of algorithms that may somehow seem new will actually
find that they also already fit inside of our geometrical framework.
Finally, we hope that readers will see that the many algorithms that
have been proposed over the past several decades are not just vaguely
connected to each other, but are elements of a deeper mathematical
structure.  The reader who sees the depth and simplicity of \S
\ref{sec:optim}, say, has understood our message.

\subsection{Rayleigh Quotient Iteration}
\label{sec:rqi}

If $A$ is a symmetric matrix, it is well known that Rayleigh quotient
iteration (RQI) is a cubically convergent algorithm.  It is easy to
derive formulas and show that it is true, here we will explain our
view of {\em why\/} it is true.  Let $r(x)$ denote the Rayleigh
quotient $x\T Ax$ and, abusing notation, let $r(\theta)$ denote the
Rayleigh quotient on a geodesic with $\theta=0$ corresponding to an
eigenvector of $A$.

Here is the intuition.  Without writing down any formulas, it is
obvious that $r(\theta)$ is an even function of $\theta$; hence
$\theta=0$ is an extreme point.  Newton's optimization method, usually
quadratically convergent, converges cubically on nondegenerate even
functions.  Keeping in mind that $A-r(x)I$ is the second covariant
derivative of the Rayleigh quotient, inverting it must amount to
applying Newton's method.  Following this intuition, RQI must converge
cubically. The intution is that simple.

Indeed, along a geodesic, $r(\theta)=\lambda \cos^2\theta +\alpha
\sin^2\theta$ (we ignore the degenerate case $\alpha=\lambda$).
The $k$th step of Newton's method for the univariate function
$r(\theta)$ is readily verified to be
$$\theta_{k+1}=\theta_k- \half\tan(2 \theta_k) =
-{\textstyle\frac{4}{3}}\theta_k^3 + O(\theta_k^5).$$ We think of
updating $\theta$ as moving along the circle.  If we actually moved
tangent to the circle by the Newton update $-\half\tan(2 \theta_k)$
and then projected to the circle, we would have the Rayleigh quotient
iteration $$\theta_{k+1}=\theta_k- \arctan\bigl(\half\tan
(2\theta_k)\bigr) = -\theta_k^3 + O(\theta_k^5).$$ This is the
mechanism that underlies Rayleigh quotient iteration.  It ``thinks''
Newton along the geodesic, but moves along the tangent. The angle from
the eigenvector goes from $\theta$ to $-\theta^3$ almost always.
(Readers comparing with Parlett \cite[Eq.~(4-7-3)]{parlett80} will
note that only positive angles are allowed in his formulation.)

\begin{figure}[t]
\centerline{\epsfbox[20 552 240 775]{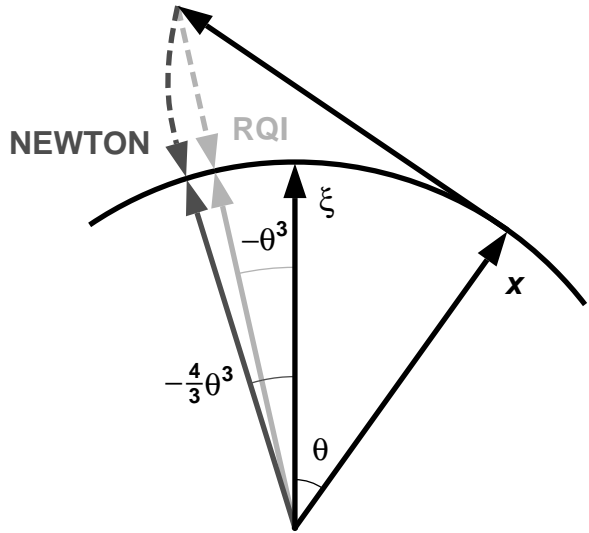}}
\caption{Cubic convergence of RQI and Newton's method applied to Rayleigh's
quotient.  The vector $\xi$ is an eigenvector.}
\label{fig:rqic}
\end{figure}

When discussing the mechanism, we only needed one variable: $\theta$.
This is how the mechanism should be viewed because it is independent
of the matrix, eigenvalues, and eigenvectors.  The algorithm, however,
takes place in $x$ space.  Since $A-r(x)I$ is the second covariant
derivative of $r(x)$ in the tangent space at $x$, the Newton update
$\delta$ is obtained by solving $\Pi (A-r(x)I)\delta =-\Pi Ax=
-(A-r(x)I)x$, where $\Pi=I-xx^T$ is the projector.  The solution is
$\delta=-x+y/(x^Ty)$, where $y=(A-r(x)I)^{-1}x$.  The Newton step
along the tangent direction is then $x \rightarrow x+\delta=
y/(x^Ty)$, which we project to the unit sphere.  This is exactly an
RQI step.  These ideas are illustrated in Figure~\ref{fig:rqic}.

One subtlety remains.  The geodesic in the previous paragraph is
determined by $x$ and the gradient rather than $x$ and the
eigenvector.  The two geodesics converge to each other by the inverse
iteration process (almost always) allowing the underlying mechanism to
drive the algorithm.
\goodbreak

One trivial example where these issues arise is the generalization and
derivation of Davidson's method \cite{vorst96,davidson75,sadkane94}.
In this context there is some question as to the interpretation of
$D-\lambda I$ as a preconditioner.  One interpretation is that it
preconditions the eigenproblem by creating better eigenvalue spacings.
We believe that there is a more appropriate point of view.  In linear
conjugate gradient for $Ax=b$, preconditioners are used to invert $M$
which is an approximation to $A$ (the Hessian of $\frac{1}{2}x\T
Ax-x^Tb$) against the gradient.  This is an approximate Newton
step. In nonlinear conjugate gradient, there is no consensus as to
whether inverting the Hessian (which is approximated by $D-\lambda
I$!)  would constitute the ideal preconditioner, but it is a Newton
step.  Therefore with the link between nonlinear conjugate gradient
preconditioning and approximate Newton step, we see that Davidson's
method is deserving of being called a preconditioner from the
conjugate gradient point of view.

\subsection{Coordinate Singularities of Symmetric Matrices}
\label{sec:coord}

An important open problem in numerical linear algebra is the complete
understanding of the influence of singularities on computations
\cite{demmelkahan,chatelin96}.  In this section we shall describe the
singularity associated with multiple eigenvalues of symmetric matrices
in terms of coordinate singularities, i.e., the breakdown of the
coordinate representation.  In \S{}\ref{sec:optim}, we will
describe how understanding this coordinate singularity underlies a
regularization approach to eigenvalue optimization.

Matrix factorizations are nothing more than changes in variables or
coordinate changes.  In the plane, Cartesian and polar coordinates
both give orthogonal systems, but polar coordinates have a coordinate
singularity at the origin.  A small perturbation near the origin can
violently change the angle coordinate.  This is ill-conditioning.  If
the $r$ coordinate goes through the origin we have a singularity of
the form $|r|$.

Consider traceless, symmetric, $2 \by 2$ matrices: $$ A=
\pmatrix{x&y\cr y&-x\cr}.$$ The positive
eigenvalue is $r=\sqrt{\strut x^2+y^2}$, and one of the orthogonal
eigenvectors is ${\cos{1\over2}\theta\atopwithdelims()
\sin{1\over2}\theta}$, where $\tan \theta=y/x$.  The
conversion between matrix elements and the eigendecomposition is
exactly the conversion from Cartesian to polar coordinates.  Whatever
ill-conditioning one associates with a symmetric matrix with two close
eigenvalues, it is the same numerical difficulty associated with the
origin in polar coordinates.  The larger eigenvalue behaves like $|r|$
at the origin, and the eigenvector behaves like $\theta$ changing
violently when perturbed.  If one wants to think about all $2
\by 2$ symmetric matrices, add $z$ as the trace, and the resulting
interpretation is cylindrical coordinates.

We now generalize.  Let $S_n$ be the space of $n \by n$ symmetric
matrices.  Suppose that the largest $p$ eigenvalues
$\lambda_1,\ldots,\lambda_p$ coalesce.  The corresponding eigenvectors
are not uniquely determined, but the invariant subspace is.
Convenient parameterizations are
$$
\vbox{\halign{\hfil$\displaystyle#$&\hfil$\displaystyle{}#{}$\hfil
&\hfil#\hfil&\hfil$\displaystyle{}#{}$\hfil &\hfil$\displaystyle#$\cr
S_n&\equiv&Symmetric Matrices&=&\R^p \times V_{n,p}\times S_{n-p}\cr
S_{n,p}&\equiv&$\{\,S_n:\hbox{$\lambda_1$ has multiplicity
$p$}\,\}$&=&\R\times G_{n,p} \times S_{n-p}\cr}}
$$
\goodbreak
\noindent 
That is, any
symmetric matrix may be parameterized by its $p$ largest eigenvalues,
the corresponding eigenvectors in order, and the $(n-p)\by(n-p)$
symmetric operator on the space orthogonal to these eigenvectors.  To
parameterize a symmetric matrix with eigenvalue $\lambda$ of
multiplicity $p$, we must specify the invariant subspace corresponding
to this eigenvalue and, once again, the $(n-p)\by(n-p)$ symmetric
operator on the orthogonal subspace.  It is worth mentioning that the
parameters in these decompositions give an orthonormal system
(surfaces with constant parameters intersect orthogonally).  The
codimension of $S_{n,p}$ in $S_n$ is $p(p+1)/2-1$, obtained by adding
$p-1$ (corresponding to $\lambda_2$, \dots, $\lambda_p$) to $p(p-1)/2$
(the codimension of $G_{n,p}$ in $V_{n,p}$).

Another interpretation of the well-known fact that when eigenvalues
coalesce, eigenvectors, but not invariant subspaces, are
ill-conditioned, is that the Stiefel manifold collapses to the
Grassmann manifold.  As with polar coordinates we have a coordinate
singularity corresponding to ill-conditioning near $S_{n,p}$.  Near
this set, a small perturbation will violently move the Stiefel
component.  
The singularity associated with the coallescing of eigenvalues
is very much the singularity of the function $f(x)=|x|$.

\subsection{The CS Decomposition}
\label{sec:csdecomp}

The CS decomposition \cite{golub89a} should be recognized as the
geodesic between two points on the Grassmann manifold.  Any $n\by n$
orthogonal matrix $Q$ may be written as
\begin{equation} Q =\pmatrix{V&0\cr 0&U\cr} \pmatrix{C&-S&0\cr
S&C&0\cr 0&0&I\cr} \pmatrix{\tilde V&0\cr 0&\tilde U\cr}^T 
\end{equation} for
some $p\by p$ orthogonal matrices $V$ and $\tilde V$ and
$(n-p)\by(n-p)$ orthogonal matrices $U$ and $\tilde U$, and $p$
angles $\theta_i$ where
$C=\diag(\cos\theta_1,\ldots,\cos\theta_p)$ and
$S=\diag(\sin\theta_1,\ldots,\sin\theta_p)$.  Comparing this with
the geodesic formula Eq.~(\ref{eq:csgeos}) and letting
$\theta_i=t\sigma_i$ ($i=1$, \dots, $p$) where $\sigma_i$ are the
diagonal elements of~$\Sigma$, we see that the first $p$ columns
of the CS decomposition traverse a geodesic emanating from
$Y(0)={I\atopwithdelims()0}$ (the origin).  The next $p$ columns
give an orthogonal basis for the velocity vector along the
geodesic (in fact, they are the orthogonal component of its polar
decomposition).

As is well known, the $\theta_i$ are the principal angles between the
subspaces spanned by the first $p$ columns of~$Q$ and the origin.  In
general, let $\theta_i$ ($i=1$, \dots, $p$) be the principal angles
between the two subspaces spanned by the columns of $n\by p$
orthonormal matrices $Y_1$ and $Y_2$, i.e., $U(\cos\Theta)V^T$ is the
singular value decomposition of~$Y_1^TY_2$, where $\Theta$ is the
diagonal matrix of principal angles.  Also let $\theta$ and
$\sin\theta$ represent the $p$-vectors formed by the $\theta_i$ and
$\sin\theta_i$.  These principal angles provide several different
definitions of the distance between two subspaces:
\goodbreak

\begin{enumerate}
\itemsep=3pt
\item {\bf arc~length:}\hskip1em minus.5em
   ${d(Y_1,Y_2) =\|\theta\|_2}$
\item {\bf Fubini-Study:}\hskip1em minus.5em
   ${d_{\rm FS}(Y_1,Y_2)=\arccos|\det Y_1^TY_2|=\arccos(\prod_i\cos\theta_i)}$
\item {\bf chordal $2$-norm:}\hskip1em minus.5em ${d_{c2}(Y_1,Y_2)
   =\|Y_1U-Y_2V\|_2 =\|2\sin\half\theta\|_\infty}$
\item {\bf chordal Frobenius-norm:}\hskip1em minus.5em
   ${d_{cF}(Y_1,Y_2) =\|Y_1U-Y_2V\|_F =\|2\sin\half\theta\|_2}$
\item {\bf projection 2-norm
    {\rm\cite{golub89a}}:}\hskip1em minus.5em 
   ${d_{p2}(Y_1,Y_2) =\|Y_1Y_1^T-Y_2Y_2^T\|_2 =\|\sin\theta\|_\infty}$
\item {\bf projection F-norm:}\hskip1em minus.5em
   ${d_{pF}(Y_1,Y_2) =2^{-1/2}\|Y_1Y_1^T  - Y_2Y_2^T\|_F=\|\sin\theta\|_2}$
\end{enumerate}
\goodbreak

The arc~length distance is derived from the intrinsic geometry of the
Grassmann manifold.  The chordal $2$-norm and Frobenius-norm distances
are derived by embedding the Grassmann manifold in the vector space
$\R^{n\times p}$, then using the the $2$- and Frobenius-norms,
respectively, in these spaces.  Note that these distances may be
obtained from the minimization problems $$d_{c2\ {\rm or}\
cF}(Y_1,Y_2)= \min\limits_{Q_1,Q_2\in O_p}\|Y_1Q_1-Y_2Q_2\|_{2\ {\rm
or}\ F}.$$ The projection matrix $2$-norm and Frobenius-norm distances
are derived by embedding the Grassmann manifold in the set of $n\by n$
projection matrices of rank~$p$, then using the $2$- and
Frobenius-norms, respectively.  The Fubini-Study distance is derived
via the Pl\"ucker embedding of~$G_{n,p}$ into the projective space
$\PS(\bigwedge^p(\R^n))$ (by taking wedge products between all columns
of~$Y$), then using the Fubini-Study metric
\cite{kobayashi}.\footnote{We thank Keith Forsythe for reminding us of
this distance.} Note that all metrics except the chordal and
projection matrix $2$-norm distances are asymptotically equivalent for
small principal angles, i.e., these embeddings are isometries, and that
for $Y_1\ne Y_2$ we have the strict inequalities
$$
\displaylines{\refstepcounter{equation}\hfill d(Y_1,Y_2)> d_{\rm
FS}(Y_1,Y_2), \hfill\llap{(\theequation)}\cr
\refstepcounter{equation}\hfill d(Y_1,Y_2)> d_{cF}(Y_1,Y_2)>
d_{pF}(Y_1,Y_2), \hfill\llap{(\theequation)}\cr
\refstepcounter{equation}\hfill d(Y_1,Y_2)> d_{cF}(Y_1,Y_2)>
d_{c2}(Y_1,Y_2), \hfill\llap{(\theequation)}\cr
\refstepcounter{equation}\hfill d(Y_1,Y_2)> d_{pF}(Y_1,Y_2)>
d_{p2}(Y_1,Y_2). \hfill\llap{(\theequation)}\cr}
$$
\goodbreak\noindent 
These inequalities
are intuitively appealing because by embedding the Grassmann manifold
in a higher dimensional space, we may ``cut corners'' in measuring the
distance between any two points.

\goodbreak
\subsection{Conjugate Gradient for the Eigenvalue Problem}
\label{sec:cgeig}

Conjugate gradient algorithms to minimize $\frac{1}{2}y^T\! Ay$
($A$ symmetric) on the sphere ($p=1$) is easy and has been
proposed in many sources.  The correct model algorithm for \hbox{$p>1$}
presented in this paper is new.  We were at first bewildered by
the number of variations
\cite{alsen71a,bradbury66a,foxkapoor69,fried69,andersson71,geradin71a,fried72,fu95,ruhe74,%
sameh82a,fuhrmann84,perdon86,chen86,cohen87,yang89}
most of which propose ``new'' algorithms for conjugate gradient
for the eigenvalue problem.  Most of these algorithms are for
computing extreme eigenvalues and corresponding eigenvectors.  It
is important to note that none of these methods are equivalent to
Lanczos \cite{edelman96a}.  It seems that the correct approach to
the conjugate gradient algorithm for invariant subspaces ($p >1$)
has been more elusive.  We are only aware of three papers
\cite{alsen71a,sameh82a,fu95} that directly consider conjugate
gradient style algorithms for invariant subspaces of dimension
\hbox{$p>1$.}  None of the proposed algorithms are quite as close to the
new idealized algorithms as the $p=1$ algorithms are.  Each is
missing important features which are best understood in the
framework that we have developed.  We discuss these algorithms
below.

The simplest non-trivial objective function on the Grassmann manifold
$G_{n,p}$ is the quadratic form $$F(Y)=\ha \tr Y\T A Y,$$ where $A$ is
a symmetric $n\by n$ matrix.  It is well known that the solution
to the minimization of $F$ is the sum of the $p$ smallest eigenvalues
of $A$, with an optimal $Y$ providing a basis for the invariant
subspace corresponding to the $p$ smallest eigenvalues.

To solve the eigenvalue problem, one may use the template directly
from \S{}\ref{sec:cggrass} after deriving the gradient
$$\nabla F(Y)=AY-Y(Y\T AY)$$ and the second covariant derivative of
$F(Y)$: $${\Hess F} (\Delta_1,\Delta_2) =
\tr\bigl(\Delta_1\T A\Delta_2 - (\Delta_1^T\Delta_2)Y\T AY\bigr).$$
 
The line minimization problem may be solved as $p$ separate two-by-two
problems in parallel, or it may be solved more completely by solving
the $2p\by 2p$ eigenvalue problem.  This does not follow the geodesic
directly, but captures the main idea of the block Lanczos algorithm
which in some sense is optimal \cite{cullum78,cullum85}.

If one is really considering the pure linear symmetric eigenvalue
problem then pure conjugate gradient style procedures must be inferior
to Lanczos.  Every step of all proposed non-preconditioned conjugate
gradient algorithms builds vectors inside the same Krylov space in
which Lanczos gives an optimal solution.  However, exploring conjugate
gradient is worthwhile.  When the eigenvalue problem is non-linear or
the matrix changes with time, the Lanczos procedure is problematic
because it stubbornly remembers past information that perhaps it would
do well to forget.  (Linear conjugate gradient, by contrast, benefits
from the memory of this past information.)  Applications towards
non-linear eigenvalue problems or problems that change in time drive
us to consider the conjugate gradient method.  Even the eigenvalue
problem still plays a worthy role: it is the ideal model problem that
allows us to understand the procedure much the way the Poisson
equation on the grid is the model problem for many linear equation
solvers.

Conjugate gradient on the sphere ($p=1$) computes the smallest
eigenvalue of a symmetric matrix $A$.  Two papers
\cite{perdon86,chen86} consider imposing conjugacy through $A$.
This is an unfortunate choice by itself because $A$ is quite
different from the Hessian $A-r(x)I$, where $r(x)$ is the
Rayleigh quotient.  A few authors directly consider conjugacy
through the unconstrained Hessian \cite{geradin71a,yang89}.
Others attempt to approximate conjugacy through the Hessian by
using Polak-Ribi\'ere or Fletcher-Reeves
\cite{bradbury66a,foxkapoor69,fried69,andersson71,fried72,fuhrmann84,cohen87,yang89,ruhe74}.
It is quite possible that most of these variations might well be
competitive with each other and also our idealized algorithm, but
we have not performed the numerical experiments because
ultimately the $p=1$ case is so trivial.  A comparison that may
be of more interest is the comparison with restarted Lanczos.  We
performed an informal numerical experiment that showed that the
conjugate gradient method is always superior to two step Lanczos
with restarts (as it should be since this is equivalent to the
steepest descent method), but is typically slightly slower than
four step Lanczos.  Further experimentation may be needed in
practice.

Turning to the $p>1$ case, the three papers that we are aware of are
\cite{alsen71a,sameh82a,fu95}.  The algorithm proposed in Als\'{e}n
\cite{alsen71a}, has a built-in extra feature not in the
idealized algorithm.  Though this may not be obvious, it has one step
of orthogonal iteration built in.  This may be viewed as a
preconditioning procedure giving the algorithm an advantage. The
Sameh-Wisniewski \cite{sameh82a} algorithm begins with many of the
ideas of an idealized Grassmann algorithm, including the recognition
of the correct tangent on the Grassmann manifold (though they only
mention imposing the Stiefel constraint). Informal experiments did not
reveal this algorithm to be competitive, but further experimentation
might be appropriate.  The more recent Fu and Dowling algorithm
\cite{fu95} imposes conjugacy through $A$ and therefore we do not
expect it to be competitive.
\goodbreak

\subsection{Conjugate Gradient for the  Generalized Eigenvalue Problem}

It is well known that the generalized eigenvalue problem $Ax=\lambda
Bx$ may also be posed as a constrained optimization problem.
Now we must find $$\min \tr Y\T AY$$ subject to the constraint that
$$Y\T BY=I_p.$$

With the change of variables
\begin{eqnarray}
&&\bar Y=B^{1/2}Y \\
&&\bar\Delta=B^{1/2} \Delta \\
&&\bar A=B^{-1/2}AB^{-1/2}
\end{eqnarray}
the problem becomes
$$\min \tr \bar Y\T\bar A\bar Y\quad \hbox{subject to}\quad \bar
Y^T\bar Y=I_p.$$ The numerical algorithm will be performed on the
non-overlined variables, but the algorithm will be mathematically
equivalent to one performed on the overlined variables.

Notice that the condition on tangents in this new coordinate system is
that $$\Delta\T BY=0.$$ It is readily checked that the gradient of the
trace minimization problem becomes $$G=(B^{-1}-YY^T)AY$$ (note that
$G\T BY=0$).


Geodesics may be followed in any direction $\Delta$ for which
$\Delta\T BY=0$ by computing a  compact  variation on the  SVD of $\Delta$:
$$\Delta=U\Sigma V^T,\quad \hbox{where $U\T BU=I$.}$$

For simplicity, let us assume that $\Delta$ has full rank~$p$.  The
$V$ vectors are the eigenvectors of the matrix $\Delta^TB\Delta$,
while the $U$ vectors are the eigenvectors of the matrix
$\Delta\Delta^TB$ corresponding to the non-zero eigenvalues.  There is
also a version involving the two matrices $$\pmatrix{0&0&\Delta\cr
B&0&0\cr0&\Delta^T&0\cr} \quad\hbox{and}\quad
\pmatrix{0&0&B\cr\Delta^T&0&0\cr0&\Delta&0\cr}.$$ This SVD may be
expressed in terms of the quotient SVD \cite{golub89a,demoor95}.

Given the SVD, we may follow geodesics by computing
$$Y(t)=\pmatrix{YV&U\cr}\pmatrix{C\cr S\cr} V^T.$$
All the $Y$ along this curve have the property that $Y\T BY=I$.  For
the problem of minimizing $\ha \tr Y\T AY$, line minimization
decouples into $p$ two-by-two problems just as in the ordinary
eigenvalue problem.

Parallel transport, conjugacy, and the second covariant derivative may
all be readily worked out.

\subsection{Electronic Structures Computations}

In this section, we briefly survey a research area where conjugate
gradient minimization of non-quadratic but smooth functions on the
Stiefel and Grassmann manifolds arise, the {\em ab initio\/}
calculation of electronic structure within the local density
approximation.  Such approaches use only the charge and mass of
electrons and atomic nuclei as input and have greatly furthered
understanding of the thermodynamic properties of bulk materials
\cite{buda90}, the structure and dynamics of surfaces
\cite{rsurfdyn,rGesur}, the nature of point defects in crystals
\cite{rosi}, and the diffusion and interaction of impurities in bulk
materials \cite{rDiff}.  Less than ten years ago, Car and Parrinello
\cite{car} in a watershed paper proposed minimization through
simulated annealing.  Teter and Gillan
\cite{gillan89a,teter89a} later introduced conjugate gradient based
schemes and demonstrated an order of magnitude increase in the
convergence rate.  These initial approaches, however, ignored entirely
the effects of curvature on the choice of conjugate search directions.
Taking the curvature into {\em partial\/} account using a
generalization of the Riemannian projection led to a further
improvement in computation times by over a factor of three under
certain conditions \cite{arias92a}.

Our ability to compute {\em ab initio}, using only the charge and mass
of electrons and atomic nuclei as input, the behavior of systems of
everyday matter has advanced greatly in recent years.  However, the
computational demands of the approach and the attendant bounds on the
size of systems which may be studied (several hundred atoms) have
limited the direct impact of the approach on materials and chemical
engineering.  Several {\em ab initio\/} applications which will
benefit technology tremendously remain out of reach, requiring an
order of magnitude increase in the size of addressable systems.
Problems requiring the simultaneous study of thousands of atoms
include defects in glasses (fiber optics communications), complexes of
extended crystalline defects (materials' strength and processing), and
large molecules (drug design).

The theoretical problem of interest is to find the smallest eigenvalue
$E_0$ of the Schr\"{o}dinger equation in the space of $3N$ dimensional
skew-symmetric functions, $$H\psi=E_0 \psi,$$ where the Hamiltonian
operator $H$ is defined by $$ H = \sum_{1\le n\le N}{\left(
-\frac{1}{2}\nabla_n^2+V_{ion}(r_n)\right)}+\frac{1}{2}\sum_ {1<n\ll m
\le N}{\frac{1}{\|r_n-r_m\|^2}}.$$ Here, $N$ is the number of
electrons in the system under study, now typically on the order of
several hundred, $r_i$ is the position of the $i$th electron,
$V_{ion}(r)$ is the potential function due to the nuclei and inner
electrons, and the second summation is recognized as the usual Coulomb
interactions. Directly discretizing this equation at $M$ grid-points
in space would lead to absurdly huge eigenvalue problems where the
matrix would be $M^N \by M^N$.  This is not just a question of dense
versus sparse methods, a direct approach is simply infeasible.

The fundamental theorems which make the {\em ab initio\/} approach
tractable come from the density functional theory of Hohenberg and
Kohn \cite{hohenberg} and Kohn and Sham \cite{kohn}.  Density
functional theory states that the ground states energy of a quantum
mechanical system of interacting electrons and ions is equal to the
solution of the problem of minimizing an energy function over all
possible sets of $N$ three-dimensional functions (electronic
orbitals) obeying the constraints of orthonormality.  Practical
calculations generally use a finite basis to expand the orbitals, but
for purposes of discussion, we may discretize the problem onto a
finite spatial grid consisting of $M$ points.  The Kohn-Sham
minimization then becomes,
\begin{eqnarray} \label{eqn:E_KS}
E_0 & = & \min_{X^TX=I_N} E(X) \\
& \equiv & \min_{X^TX=I_N} { \tr(X^T \! H X)+f\bigl(\rho(X)\bigr)}, \nonumber
\end{eqnarray}
where each column of $X$ is a different electronic orbital sampled on
the spatial grid, $\rho$ is the vector $\rho_i(X)\equiv\sum_n
|X_{in}|^2$, $H$ is an $M \by M$ matrix (single-particle Hamiltonian),
and $f$ is a function which we leave unspecified in this discussion.
In full generality the $X$ are complex, but the real case applies for
physical systems of large extent that we envisage for this application
\cite{ptaaj}, and we, accordingly, take $X$ to be real in this discussion.

Recent advances in computers have enabled such calculations on systems
with several hundreds of atoms \cite{screw,brommer}.  Further
improvements in memory and performance will soon make feasible
computations with upwards of a thousand atoms.  However, with growing
interest in calculations involving larger systems has come the
awareness that as the physical length of systems under study
increases, the Hessian about the minimum of Eq.~(\ref{eqn:E_KS}) becomes
increasingly ill-conditioned and non-conjugate minimization approaches
exhibit a critical slowing down \cite{teter89a}.  This observation
prompted workers \cite{gillan89a,teter89a} to apply conjugate gradient
concepts to the problem, and now dozens of researchers have written papers
using some form of the conjugate gradient method.  In particular, one
has a Grassmann problem when the number of electrons in each state is
constant (i.e., two one spin up and one spin down).  This is what
happens in calculations on semiconductors and ``closed shell'' atoms
and molecules.  Otherwise, one has a Stiefel problem such as when one
has metals or molecules with partially filled degenerate states.

The framework laid out in this discussion may be of practical use to the
{\em ab initio\/} density-functional community when the inner product
computation through the Hessian of $E(X)$ is no more computationally
complex to evaluate than calculating the energy function $E(X)$ or
maintaining the orthonormality constraints $X^T X = I_N$.  A suitable
form for this inner product computation is
\begin{eqnarray} \label{eqn:Hess_over}
\frac{1}{2} \sum_{in,jm} {Y_{in}
\frac{\partial^2 E}{\partial X_{in} \partial X_{jm}} Z_{jm}} & = &
\tr\bigl(Y^T(H+V)Z\bigr) + \sum_{ij}{  \sigma_i \left(2
\frac{\partial^2\!f}{\partial \rho_i
\partial \rho_j}\right) \tau_j} \\
&&\quad{} - \tr\bigl(X^T (H+V) (X Y^T Z)\bigr) \nonumber
\end{eqnarray}
\goodbreak
\noindent
where $V$ is the diagonal matrix defined by
$V_{ij}=(\partial f/\partial \rho_i) \delta_{ij}$, $\sigma_i \equiv
\sum_n{Y_{in} X_{in}}$, $\tau_i \equiv
\sum_n{Z_{in} X_{in}}$.
Written this way, the first two terms of Eq.~(\ref{eqn:Hess_over}) have
the same form and may be evaluated in the same manner as the
corresponding terms in Eq.~(\ref{eqn:E_KS}), with $\sigma$ and $\tau$
playing roles similar to $\rho$.  The third term, coming from the
curvature, may be evaluated in the same way as the first term of
Eq.~(\ref{eqn:Hess_over}) once given the object $XY^TZ$, which is no
more computationally complex to obtain than the Gram-Schmidt
orthonormalization of an object like $X$.

\subsection{Subspace Tracking}

The problem of computing the principal invariant subspace of a
symmetric or Hermitian matrix arises frequently in signal processing
applications, such as adaptive filtering and direction finding
\cite{owsley78,schmidt79,bienvenu83,schreiber86,roy89}.  Frequently,
there is some time-varying aspect to the signal processing problem,
and a family of time-varying principal invariant subspaces must be
tracked.  The variations may be due to either the addition of new
information as in covariance matrix updating, a changing signal
environment, or both.  For example, compute the principal invariant
subspace of either of the covariance matrices \begin{eqnarray}
R_k&=&R_{k-1} +x_kx_k^T\qquad \hbox{$k=1$, $2$, \dots, and $x_k$ is
given} \label{eq:r1upd}\\ R(t)&=&\hbox{a continuous function of~$t$}
\label{eq:Rcont}\end{eqnarray} at every iteration or at discrete times.
Eq.~(\ref{eq:r1upd}) typically arises from updating the sample
covariance matrix estimate; Eq.~(\ref{eq:Rcont}), the more general
case, arises from a time-varying interference scenario, e.g.,
interference for airborne surveillance radar \cite{ward94,smith95}.
Solving this eigenvalue problem via the eigenvalue or singular value
decompositions requires a large computational effort. Furthermore,
only the span of the first few principal eigenvectors may be required,
whereas decomposition techniques compute all eigenvectors and
eigenvalues, resulting in superfluous computations.  Approaches to
this problem may be classified as standard iterative methods
\cite{golub89}, methods exploiting rank~1 updates
\cite{owsley78,karhunen84,schreiber86,yu91,moonen92,stewart92,champagne94,mathew95},
i.e., Eq.~(\ref{eq:r1upd}), Lanczos based methods
\cite{comon90,xu94a,xu94b}, gradient based methods
\cite{owsley78,yang88,brockett91b}, conjugate gradient based methods
\cite{fuhrmann84,chen86,sarkar89,yang89,smith93a,fu95,smith96a}, which
are surveyed by Edelman and Smith \cite{edelman96a}, Rayleigh-Ritz
based methods \cite{fuhrmann88,comon90}, and methods that exploit
covariance matrix or array structure \cite{roy89,xu94a,xu94b}.

If the subspace does not change quickly over (discrete or continuous)
time, then the desired solution will be close to the previously
computed solution, and an iterative gradient-based algorithm such as
the conjugate gradient algorithm may be computationally attractive for
the subspace tracking problem.  Thus the subspace tracking problem is
treated as a time-varying optimization problem.  Other conjugate
gradient methods for computing principal invariant subspaces in a
signal processing context have appeared
\cite{chen86,sarkar89,yang89,fu95}; however, these conjugate gradient
techniques do not exploit the structure of the subspace constraint
(see \S{}\ref{sec:cgeig}).  Instead, we employ the conjugate
gradient method on the Grassmann manifold, or an approximation of it
discussed in \S{}\ref{sec:numres}.  Comon and Golub \cite{comon90}
describe and compare a wide variety of different algorithms for the
problem of exponential covariance matrix updates, with particular
emphasis on Lanczos and gradient-based algorithms. Yang, Sarkar, and
Arvas \cite{yang89} survey some conjugate gradient algorithms applied
to computing the principal invariant subspace of a fixed symmetric
matrix.  We adopt the general assumption that the matrix may change
arbitrarily over time, but that it must vary ``slowly enough'' so that
using a conjugate gradient based approach is computationally
efficient.  This last constraint is, of course, dependent upon the
application.  For the example of\break
\goodbreak
\noindent
space-time adaptive processing for
airborne radar with a rotating antenna, Smith~\cite{smith96a} shows
that this method is capable of tracking the principal invariant
subspace of clutter interference; however, when the interference
dimension $p$ is increased to account for new interference
eigenvalues, one does better to compute the eigendecomposition from
scratch and use it to initiate a new subspace track.

\subsection{Newton's Method for Invariant Subspace Computations}
\label{sec:nmisc}

Methods for refining estimates for invariant subspace computations
have been proposed by Chatelin \cite{chatelin84,chatelin93}, Dongarra,
Moler, and Wilkinson \cite{dongarra83}, and Stewart~\cite{stewart73}.
Demmel \cite[\S{}3]{demmel87} proposes a unified approach by
showing that they are all solutions to a Riccati equation.

These algorithms, when applied to symmetric matrices, are all
variations on our geometrical Newton algorithm and may be understood
in this context.  There is nothing special about the eigenvalue
problem; Newton's method for any function on the Grassmann manifold
yields a Sylvester equation in the tangent space.  The reason a
Riccati equation arises rather than a Sylvester equation is that the
previous algorithms formulate the problem in an affine space with
arbitrary constraints.  Previous researchers knew the quadratic term
in the Riccati equation belonged there, and knew that it somehow is
related to the orthogonality constraints, but we now see that it is an
artifact of a flat space derivation.

Let us take a closer look.  Previous researchers proposed algorithms
for invariant subspaces by asking for a solution to the matrix
equation $$AY-YB=0$$ made nondegenerate by imposing the affine
constraint $$Z^TY=I,$$ for some arbitrary choice of $Z$.  In the
Dongarra et al.\ case, $Z$ may be obtained by inverting and
transposing an arbitrary $p \times p$ minor of the $n
\times p$ matrix $Y$.  In Moler's Matlab notation 
\verb+Z=zeros(n,p); Z(r,:)=inv(Y(r,:))'+, 
where \verb+r+ 
denotes a $p$-vector of row
indices.  For Stewart, $Z=Y(Y^TY)^{-1}$.

A mathematically insightful approach would require no arbitrary choice
for $Z$.  We would simply specify the problem by performing Newton's
method on the function $F(Y)=\ha \tr Y\T A Y$ on the Grassmann
manifold.  The stationary points of $F(Y)$ are the invariant
subspaces.  There is no need to specify any further constraints and
there are no degeneracies.  (Notice that asking for the solution to
$AY=Y(Y\T AY)$ subject to $Y^TY=I$ is a degenerate problem.)

Newton's method requires the solution $\Delta$ to the Sylvester
equation $$\Pi\bigl(A\Delta -\Delta(Y\T AY)\bigr)=-\Pi AY,$$ where
$\Pi=(I-YY^T)$ denotes the projection onto the tangent space of the
Grassmann manifold and $G=\Pi AY$ is the gradient.  The solution is
$\Delta=-Y+Z(Y^TZ)^{-1}$, where $Z$ is the solution to the Sylvester
equation $AZ-Z(Y\T AY)=Y$.  $Y$ may be chosen so that $Y\T AY$ is
diagonal, yielding simultaneous Rayleigh quotient iterations.  If we
move along the tangent and project rather than the geodesic we have
the iteration sending $Y$ to the $Q$ factor in the QR decomposition
of~$Z$.

\subsection{Reduced Gradient Methods, Sequential Quadratic Programming,
and Lagrange Multipliers}
\label{sec:lagrange}

In this section, we generalize beyond the Stiefel and Grassmann
manifolds to show how the language and understanding of differential
geometry provides insight into well-known algorithms for general
non-linear constrained optimization.  We will show the role that
geodesics play in these algorithms.  In the next subsection, we will
then apply the geometrical intuition developed here to directly
formulate regularized sequential quadratic programs as is needed in
eigenvalue optimization.

Here we study sequential quadratic programming (SQP) and reduced
gradient methods (RGM\null).  By SQP we mean the algorithm denoted as
Newton SQP by Boggs and Tolle
\cite[p.~14]{boggs95}, SQP by Nash and Sofer \cite[p.~512]{nash95},
and QP-based projected Lagrangian by Gill, Murray, and Wright
\cite[p.~238, Eq.~(6.41)]{gill81}.  By RGM, we specifically mean the
method sometimes denoted as the reduced Hessian method
\cite[p.~25]{boggs95}, other times simply denoted RGM
\cite[p.~520]{nash95}, and yet other times considered an example of an
RGM \cite[p.~221, Eq.~(6.17)]{gill81}.  The difference is that RGM is
derived based (roughly) on the assumption that one starts at a
feasible point, whereas SQP does not.

We begin by interpreting geometrically the Lagrangian function as
it is used in constrained optimization.  Consider the
optimization problem \begin{equation} \min_{x\in\R^n}
  f(x)\quad\hbox{given the constraint that }\quad h(x)=0\in\R^p.
  \label{eq:constropt} \end{equation} For simplicity we consider
the case where the level surfaces $h(x)=c$ are manifolds
($\partial h/\partial x$ has full rank everywhere) and we work
with the Euclidean metric.  In the Euclidean case, the
formulations are routine in the optimization community, but we
have not seen the geometric intuition (particularly geometric
interpretations away from the optimization point and the role
that geodesics play ``behind-the-scenes'') in the optimization
references that we have consulted.  Numerical Lagrange multiplier
issues are discussed in \cite{gill79} and \cite{gill81}, for
example.  In this paper, we give the new interpretation that the
Hessian of the Lagrangian is the correct matrix for computing
second derivatives along geodesics at every point, not only as an
approximation to the result at the optimal point.

At every point $x\in\R^n$, it is possible to project the gradient of
$f$ onto the tangent space of the level surface through $x$.  This
defines a sort of flattened vector field.  In terms of formulas,
projection onto the tangent space (known as computing least-squares
Lagrange multiplier estimates) means finding $\lambda$ that minimizes
the norm of \begin{equation} \pldx =f_x -\lambda\cdot h_x,
\label{eq:lagrmult} \end{equation} i.e., \begin{equation} 
\lambda={f_x}h_x^T (h_x h_x^T)^{-1}.
\label{eq:lagrls}\end{equation} At every point $x \in\R^n$ (not
only the optimal point) Lagrange multipliers are the coordinates
of $f_x$ in the normal space to a level surface of the
constraint, i.e., the row space of $h_x$.  (Our convention is
that $f_x$ is a $1 \by n$ row vector, and $h_x$ is a $p\by n$
matrix whose rows are the linearizations of the constraints.)

If $x(t)$ is any curve starting at $x(0)=x$ that is constrained to the
level surface at $x$, then $\pldx \dot{x}$ computes the derivative of
$f$ along the curve.  (In other words, $\pldx$ is the first covariant
derivative.)  The second derivative of $f$ along the curve is 
\begin{equation}
\label{eqlxx}\frac{d^2}{dt^2}f\bigl(x(t)\bigr)=\dot{x}^T \pldxx 
\dot{x}+\pldx \ddot{x}. \end{equation} At the optimal point
$\pldx$ is $0$, and therefore $\pldxx$ is a second order model
for $f$ on the tangent space to the level surface.  The vanishing
of the term involving $\pldx$ at the optimal point is well-known.

The idea that we have not seen in the optimization literature and that
we believe to be new is the geometrical understanding of the quantity
at a non-optimal point: At any point at all, $\pldx$ is tangent to the
level surface while $\ddot{x}(t)$ is normal when $x$ is a geodesic.
The second term in Eq.~(\ref{eqlxx}) conveniently vanishes here too
because we are differentiating along a geodesic!  Therefore, the
Hessian of the Lagrangian has a natural geometrical meaning, it is the
second derivative of $f$ along geodesics on the level surface, i.e.,
it is the second covariant derivative in the Euclidean metric.

We now describe the RG method geometrically.  Starting at a point $x$
on (or near) the constraint surface $h(x)=0$, the quadratic function
$$\pldx \dot{x} + \half \dot{x}^T\pldxx \dot{x}$$ models $f$ (up to a
constant) along geodesics emanating from $x$.  The $\dot{x}$ that
minimizes this function is the Newton step for the minimum for $f$.
Intrinsic Newton would move along the geodesic in the direction of
$\dot{x}$ a length equal to $\|\dot{x}\|$.  Extrinsically, we can move
along the tangent directly from $x$ to $x+\dot{x}$ and then solve a
set of nonlinear equations to project back to the constraint surface.
This is RGM\null.  It is a static constrained Newton method in that the
algorithm models the problem by assuming that the points satisfy the
constraints rather than trying to dynamically move from level surface
to level surface as does the SQP.

In SQP, we start on some level surface.  We now notice that the
quadratic function \begin{equation} \pldx \dot{x} +\half
  \dot{x}^T \pldxx \dot{x} \label{quap}\end{equation} can serve
as a model not only the first and second covariant derivative of
$f$ on the level surface through $x$ but also on level surfaces
for points near $x$.  The level surface through $x$ is specified
by the equation $h_x \dot{x}=0$.  Other parallel level surfaces
are $h_x \dot{x}+c=0$.  The right choice for $c$ is $h(x)$, which
is a Newton step towards the level surface $h(x)=0$.  Therefore
if the current position is $x$, and we form the problem of
minimizing $\pldx \dot{x} + \frac{1}{2} \dot{x}^T\pldxx \dot{x}$
subject to the constraint that $h_x\dot{x}+h(x)=0$, we are
minimizing our model of $f$ along geodesics through a level
surface that is our best estimate for the constraint $h(x)=0$.
This is the SQP method.

Practicalities associated with implementing these algorithms are
discussed in the aforementioned texts.  Generalizations to other
metrics (non-Euclidean) are possible, but we do not discuss this in
detail.  Instead we conclude by making clear the relationship between
Lagrange multipliers and the Christoffel symbols of differential
geometry.

To derive the geodesic equation, let $f(x)=x_k$, the $k$th coordinate
of $x$.  From Eq.~(\ref{eq:lagrls}), the Lagrange multipliers are
$h_{x_k}^T (h_xh_x^T)^{-1}$.  Since $f_{xx}=0$ we then have that the
geodesic equations are $\ddot{x}_k=\dot{x}^T\pldxxk\dot{x}$ ($k=1$,
\dots, $n$), where $\pldxxk$ denotes, $-h_{x_k}^T(h_xh_x^T)^{-1}\cdot
h_{xx}$, the Hessian of the Lagrangian function of $x_k$.  The matrix
$\Gamma^k=-\pldxxk$ is the Christoffel symbol of differential
geometry.

\subsection{Eigenvalue Optimization}
\label{sec:optim}

The geometric approach allows the formulation of sequential quadratic
programming problems when the Lagrange multiplier formalism breaks
down due to coordinate singularities.  Specifically, the geometric
insight from the previous subsection is that during the execution of a
sequential quadratic program there are three types of directions.  The
first direction is towards the constraint manifold.  SQP performs a
Newton step in that direction.  The second family of directions is
parallel to the constraint manifold.  SQP forms a quadratic
approximation to the objective function in the parallel level surface
obtained from the Newton step.  The remaining directions play no role
in an SQP and should be ignored.

Consider the problem of minimizing the largest eigenvalue of $A(x)$,
an $n\by n$ real symmetric matrix-valued function of $x \in \R^m$ when
it is known that at the minimum, exactly $p$ of the largest
eigenvalues coalesce.  Overton and Womersley \cite{overton95}
formulated SQPs for this problem using Lagrange multipliers and
sophisticated perturbation theory.  The constraint in their SQP was
that the $p$ largest eigenvalues were identical.  We will here
consider the case of $m > p(p+1)/2$.  One interesting feature that
they observed was the non-differentiability of the largest eigenvalue
at the optimum.  Following the geometry of the previous section, a new
algorithm without Lagrange multipliers may be readily devised.  There
will be no Lagrange multipliers because there will be no consideration
of the third directions mentioned above.

We will write $A$ for $A(x)$.  Let $\Lambda=Y\T AY$, where the
orthonormal columns of~$Y$ span the invariant subspace for the $p$
largest eigenvalues of~$A$, $\lambda_1$, \dots, $\lambda_p$.  We let
$F(A)=\lambda_1$ and $\cl(A) =\tr(\Lambda) =\lambda_1 +\cdots
+\lambda_p$.  Unlike the function $F(A)$, $\cl(A)$ is a differentiable
function at the optimal point.  One might have guessed that this
$\cl(A)$ was the right $\cl(A)$, but here is how one can logically
deduce it.

The trick is to rely not on the Lagrange multiplier formalism of
constraint functions, but rather on the geometry.  Geometry has the
power to replace a long complicated derivation with a short powerful
one.  Once the techniques are mastered, geometry provides the more
intuitive understanding.  There is no convenient $h(A)$ to express the
constraint of multiple eigenvalues; artificially creating one leads to
unnecessary complications due to the coordinate singularity when one
moves from the level surface $h(A)=0$ to another level surface.  The
right way to understand the coordinate singularity was described in
\S{}\ref{sec:coord}.  The direction of the Newton step must be the
first order constraint of the coallescing of the eigenvalues.  Using
the notation of \S{}\ref{sec:coord}, the parallel directions are
the tangent vectors of $S_{n,p}$.  All other directions play no role.
The natural level surfaces are thereby obtained by shifting the $p$
largest eigenvalues by a constant, and developing the orthogonal
eigenvector matrix $Q(0)$ as in Eq.~(\ref{eq:grassgeod}).

The message from \S{}\ref{sec:lagrange} is that whatever function
we are interested in, we are only interested in the component of the
gradient in the direction parallel to $S_{n,p}$.  The very
construction of a Lagrangian $\cl$ then may be viewed as the
construction of an appropriate function with the property that $\cl_x$
is parallel to the tangent vectors of $S_{n,p}$.  Of course the
tangent space to $S_{n,p}$ (see \S{}\ref{sec:coord}) includes
projection matrices of the form $\sum_{i=1}^p \alpha_i y_iy_i^T$,
where $y_i$ is the eigenvector corresponding to $\lambda_i$, only when
the $\alpha_i$ are all equal.  This corresponds to an identical shift
of these eigenvalues.  Therefore to form the correct gradient of the
objective function $F(A)=\lambda_1$ everywhere, we should replace the
true gradient, which is well known to be the spectral projector
$y_1y_1^T$, with its component in the direction $YY^T$, which is an
$S_{n,p}$ tangent vector.  Integrating, we\break
\goodbreak
\noindent
now see that the act of
forming the Lagrangian, which we now understand geometrically to mean
replacing $y_1y_1^T$ with $YY^T$ (projecting the gradient to the
surface of uniform shifts) amounts to nothing more than changing the
objective function from $F(x)$ to $\cl(x)= \tr(\Lambda)=\tr Y^T \! AY
$.  While one might have guessed that this was a convenient
Langrangian, we deduced it by projecting the gradient of $f(x)$ on the
tangent space of a level surface.  The components of $f(x)$ that we
removed implicitly would have contained the Lagrange multipliers, but
since these components are not well defined at the coordinate
singularity, it is of little value to be concerned with them.
 
Now we must explicitly consider the dependence of~$\cl$ on~$x$.  Our
optimization step is denoted $\Delta x$, and $\dot{A}$ and $\ddot{A}$
respectively denote $[A_x \Delta x]$ and $[A_{xx}\Delta x
\Delta x]$ (notation from \cite{overton95}).  It is easy to verify
that \begin{eqnarray} \label{ee1} \pldx&=&\tr Y\T\dot{A}Y,\\
\label{ee2} \pldxx&=&\tr(Y\T\ddot{A}Y+Y\T\dot{A} \Yd+\Yd\T
\dot{A}Y),\end{eqnarray} where $\Yd$ is the solution to
\begin{equation} \label{ee3} \Yd\Lambda
  -(I-YY^T)A\Yd=(I-YY^T)\dot{A}Y
\end{equation} that
satisfies $Y^T\Yd=0$.  The resulting sequential quadratic program over
$\Delta x$ is then \begin{equation}\min\pldx+\frac{1}{2}\pldxx,
\end{equation} subject to the
linear constraint (on $\Delta x$) that \begin{equation}
  Y^T\dot{A}Y+
\Lambda=\alpha
I,\end{equation} where the scalar $\alpha$ is arbitrary.

Let us explain all of these steps in more detail.  The allowable $\Yd$
are Grassmann directions, $Y^T\Yd = 0$.  Otherwise, we are not
parallel to the constraint surface.  Equation (\ref{ee1}) is the
derivative of $Y\T AY$.  Noting that $AY=Y\Lambda$ and $Y^T \Yd=0$,
two terms disappear.  Equation (\ref{ee2}) is trivial but we note the
problem that we do not have an explicit expression for $\Yd$, we only
have $A,Y$ and $\dot{A}$.  Fortunately, the perturbation theory for
the invariant subspace is available from Equation (\ref{ee3}).  It may
be derived by differentiating $AY=Y\Lambda$ and substituting
$\dot{\Lambda}=Y\T \dot{A} Y$.\footnote{ Alert readers may notice that
this is really the operator used in the definition of ``sep'' in
numerical linear algebra texts.  The reader really understands the
theory that we have developed in this paper if he or she now can
picture the famous ``sep'' operator as a Lie bracket with a Grassmann
tangent and is convinced that this is the ``right'' way to understand
``sep''.}  The solution to Equation (\ref{ee3}) is unique so long as
no other eigenvalue of $A$ is equal to any of
$\lambda_1,\ldots,\lambda_p$.

The linear constraint on $\Delta x$ is the one that infinitesimally
moves us to the constraint surface.  It is the condition that moves us
to a diagonal matrix.  Therefore, $\dot{\Lambda}=Y\T \dot{A} Y$ when
added to $\Lambda$ must be a scalar multiple of the identity.  This is
a linear condition on $\dot{A}$ and therefore on $\Delta x$.  The
$\alpha$ does not explicitly appear in the constraint.

\begin{figure}
\epsfxsize=5.1in\epsfbox{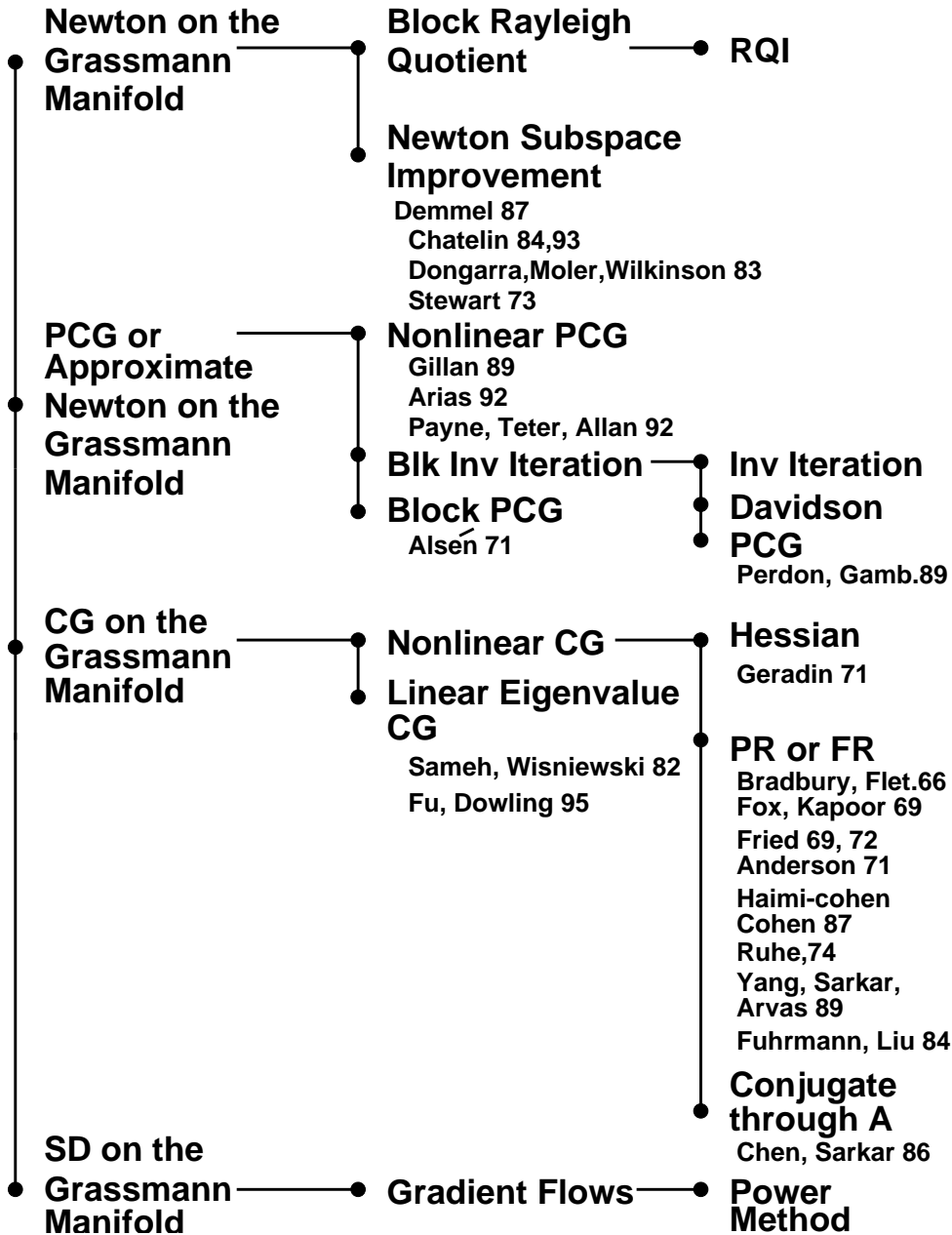}
\caption{Taxonomy of algorithms defined from the Grassmann manifold.}
\label{fig:taxonomy}
\end{figure}

\section{Conclusions}

This paper offers a new approach to the algorithms in numerical
analysis involving orthogonality constraints.  We have found that
these algorithms should be understood as optimization algorithms in
the correct geometrical setting; however, they rarely are.
\goodbreak

As a concluding example of the insight gained, we propose a Grassmann
based taxonomy for problems related to the symmetric eigenproblem.
This taxonomy allows us to view algorithms not as isolated entities,
but as objects with a coherent mathematical structure.  It is our hope
that developers of new algorithms and perturbation theories will
benefit from the analytical approach that lead to our taxonomy.

In this taxonomy, algorithms are viewed as either restrictions or
approximations of their parent.  Ultimately, we have Newton's
method on arbitrary Riemannian manifolds as the root.  One can
then restrict to a particular manifold such as the Stiefel
manifold or, as we illustrate in Figure~\ref{fig:taxonomy}, the
Grassmann manifold.  Along the vertical axis in the left column
we begin with Newton's method which may be approximated first
with PCG or approximate Newton methods, then pure conjugate
gradient, and finally steepest descent.  Moving {from} left to
right the idealized algorithms are replaced with more practical
versions that specialize for particular problems.  The second
column contains block algorithms, while the third contains single
eigenvector related algorithms.  This abstraction would not be
possible without geometry.

\section*{Acknowledgments}

The first author would like to thank Jim Demmel, Velvel Kahan, and
Beresford Parlett who have stressed the importance of geometry in
numerical linear algebra.  We also thank Scott Axelrod, Victor
Guillemin, and Shoshichi Kobayashi for a number of interesting
conversations concerning differential geometry, and Roger Brockett for
the idea of investigating conjugate gradient methods on symmetric
spaces.  We further wish to thank Dianne O'Leary, Mike Todd, Mike
Overton, and Margaret Wright for equally interesting conversations in
the area of numerical optimization.  We are indebted to the San Diego
crowd consisting of Scott Baden, Beth Ong, Ryoichi Kawai (University
of Alabama), and John Weare for working together towards understanding
the electronic structure problem.  In particular the first author
thanks John Weare for his hospitality during visits in San Diego where
we explored the issue of conjugate gradient minimization.  Steve
Vavasis asked the penetrating question of how this relates to Lagrange
multipliers which we answer in \S\ref{sec:lagrange}.
Furthermore, Ross Lippert has made a number of valuable suggestions
that are acknowledged in this paper.  We are indebted to Gene Golub
for inviting Steve Smith to the 1993 Householder Symposium in Lake
Arrowhead, California which serendipitously launched this
collaboration.

We would especially like to thank our editor Mike Overton and the
reviewers who went far beyond the call of duty with many very valuable
suggestions that enhanced the readability and relevance of this
paper.
\newpage


\makeatletter
\@itempenalty=1000
\makeatother

\bibliographystyle{siam}
\bibliography{smith}

\end{document}